\documentclass[openacc]{rstransa}

\usepackage{subfigure}
\usepackage{tikz}

\newcommand{\beq}{\begin{equation}}
\newcommand{\eeq}{\end{equation}}
\newcommand{\bseq}{\begin{subequations}}
\newcommand{\eseq}{\end{subequations}}
\newcommand{\beqn}{\begin{eqnarray}}
\newcommand{\eeqn}{\end{eqnarray}}
\newcommand{\ba}{\begin{array}}
\newcommand{\ea}{\end{array}}
\newcommand{\bct}{\begin{center}}
\newcommand{\ect}{\end{center}}
\newcommand{\btmz}{\begin{itemize}}
\newcommand{\etmz}{\end{itemize}}
\newcommand{\benum}{\begin{enumerate}}
\newcommand{\eenum}{\end{enumerate}}

\newcommand{\matbegin}{
        \left[
}
\newcommand{\matend}{
        \right]
}

\newcommand{\tbt}[4]{
  \matbegin \begin{array}{cc}
       #1 & #2 \\ #3 & #4
       \end{array} \matend }
\newcommand{\thbt}[6]{
  \matbegin \begin{array}{cc}
       #1 & #2 \\ #3 & #4 \\ #5 & #6
       \end{array} \matend }
\newcommand{\tbth}[6]{
  \matbegin \begin{array}{ccc}
       #1 & #2 & #3\\ #4 & #5 & #6
       \end{array} \matend }

\newcommand{\be}{\begin{equation}}
\newcommand{\ee}{\end{equation}}

\newcommand{\cplxs}{ C\kern -.35em \rule{0.03 em}{.7 ex}~   }

\def\complex{\hbox{C\kern -.45em \rule{0.03 em}{1.5 ex}}~}

\newcommand{\bi}{\begin{itemize}}
\newcommand{\ei}{\end{itemize}}

\newcommand{\cS}{{\cal S}}
\newcommand{\cM}{{\cal M}}
\newcommand{\cN}{{\cal N}}

\newcommand{\non}{\nonumber}
\newcommand{\ds}{\displaystyle}

\newcommand{\mrd}{\mathrm{d}}
\newcommand{\mre}{\mathrm{e}}
\newcommand{\mri}{\mathrm{i}}

\newcommand{\fvec}{{\bf f}}

\newcommand{\bu}{{\bf u}}
\newcommand{\bU}{{\bf U}}

\newcommand{\blambda}{\mbox{\boldmath$\lambda$}}
\newcommand{\bphi}{\mbox{\boldmath$\phi$}}
\newcommand{\bpsi}{\mbox{\boldmath$\psi$}}

\newcommand{\p}{\partial}

\newcommand{\tc}{\textcolor}

\title{Scaling and interaction of self-similar modes in models of high-Reynolds number wall turbulence}
\author{A S Sharma$^1$, R Moarref$^{2,3}$ and B J McKeon$^2$}
\address{$^1$ Aerodynamics and Flight Mechanics group, University of Southampton, SO17 1BJ, UK\\ $^2$ Graduate Aerospace Laboratories, California Institute of Technology, CA 91125, USA\\ $^3$ Stabilis Inc., CA 90031, USA}

\subject{fluid dynamics, wall turbulence}


\corres{Ati Sharma\\
\email{a.sharma@soton.ac.uk}}

\begin{document}

\begin{abstract}
        Previous work has established the usefulness of the resolvent operator that maps the terms nonlinear in the turbulent fluctuations to the fluctuations themselves.
    Further work has described the self-similarity of the resolvent arising from that of the mean velocity profile. The orthogonal modes provided by the {resolvent analysis describe the wall-normal coherence of the motions and inherit that self-similarity.}
        In this contribution, we present the implications of this similarity for the nonlinear interaction between {modes with different scales and wall-normal locations}. By considering the nonlinear interactions {between modes}, it is shown that much of the turbulence scaling behaviour in the logarithmic region can be determined from a single arbitrarily chosen reference plane. Thus, the geometric scaling of the modes is impressed upon the nonlinear interaction between modes. Implications of these observations on the self-sustaining mechanisms of wall turbulence, modelling and simulation are outlined.
\end{abstract}

\begin{fmtext}
\section{Introduction}

A better understanding of wall-bounded turbulent flows at high Reynolds number is essential to modelling, controlling and optimising engineering systems such as large air and water vehicles. Despite developments in high Reynolds number experiments and direct numerical simulations, several aspects of the scaling and interaction of turbulent flow structures remain unknown (see for example \cite{smimckmar11}).

\end{fmtext}

\maketitle

The resolvent analysis, introduced by \cite{mcksha10}, is a framework within which to decompose and model wall turbulence. Derived from the Navier-Stokes equations (NSE) with an assumed mean flow, it is a mathematical approach that provides a set of basis functions that are optimal in a particular sense.
The potential benefits of the approach include more efficient modelling and simulation and improved understanding of the leading physical processes in wall turbulence.

The analysis naturally leads to a decomposition into travelling waves at different wavenumbers and wavespeeds \cite{Sharma.Mezic.McKeon:2016}. Closure of the system of equations equates to knowledge of the mode coefficients, which in previous work have been found by various fitting approaches \cite{Moarref.Jovanovic.Tropp.ea:2014,Gomez.Blackburn.Rudman.ea:2016}.
Viewed from this perspective, the scaling actually observed for turbulent fluctuations must be entirely a result of the separate scaling of the resolvent modes, the interaction between the modes, and the coefficients of the modes.
Similarly, fixing the coefficients without fitting requires a proper treatment of the nonlinear interactions.

Previous work \cite{moashatromckJFM13} identified a geometric self-similarity of the resolvent operator in the logarithmic region and therefore of its leading modes.
In this paper, we derive the corresponding scaling that is induced on the quadratic nonlinearity in the NSE which governs the interaction between the modes.
The present result is therefore an important step towards a complete understanding of the scaling of turbulent fluctuations in this region. Our ultimate objective is an efficient representation of the self-sustaining mechanisms underlying wall turbulence.

In what follows, section~\ref{sec.resolvent} summarises the resolvent analysis and the pertinent linear scaling results. Section~\ref{sec.scaling-weights} presents the scaling of nonlinear interaction between modes. We conclude the paper with a discussion and summary in section~\ref{sec.conclusions}.

\section{Approach}
\label{sec.resolvent}

\subsection{The resolvent operator and its modes}
\label{sec:resmodes}

A full description of the resolvent analysis applied to wall turbulence has been given in several earlier publications \cite{mcksha10,shamck13,moashatromckJFM13}. Here, we briefly review here only the key aspects required to follow the present development.

\begin{figure}
	\begin{center}
		\includegraphics[height=2.8cm]{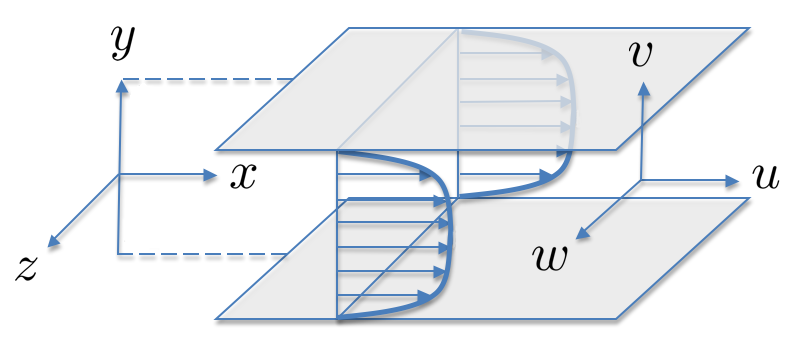}
	\end{center}
	\caption{Schematic of pressure-driven channel flow.}
	\label{fig.channel}
\end{figure}

The pressure-driven flow of an incompressible Newtonian fluid in a channel with geometry shown in figure~\ref{fig.channel} is governed by the nondimensional Navier-Stokes equations (NSE)
\begin{equation}
	\begin{array}
		{l}
		\bu_t
		\, + \,
		(\bu \cdot \nabla) \bu
		\, + \,
		\nabla P
		\; = \;
		(1/Re_\tau) \Delta \bu,
		                        \\[0.15cm]
		\nabla \cdot \bu
		\; = \;
		0,
	\end{array}
	\label{eq.NS}
\end{equation}
where $\bu (x,y,z,t) = [\,u~v~w\,]^T$ is the vector of velocities, $P (x,y,z,t)$ is the pressure, $\nabla$ is the gradient, and $\Delta = \nabla \cdot \nabla$ is the Laplacian. The streamwise, wall-normal, and spanwise directions are denoted by $x \in (-\infty,\infty)$, $y \in [0,2]$, and $z \in (-\infty,\infty)$, and $t$ denotes time. The subscript $t$ represents temporal derivative, e.g. $\bu_t = \p \bu/ \p t$. The Reynolds number $Re_\tau = u_\tau h/\nu$ is defined based on the channel half-height $h$, kinematic viscosity $\nu$, and friction velocity $u_\tau = \sqrt{\tau_w/\rho}$, where $\tau_w$ is the shear stress at the wall, and $\rho$ is the density. Unless explicitly indicated, velocity is normalized by $u_\tau$, spatial variables by $h$, time by $h/u_\tau$, and pressure by $\rho u_\tau^2$. The spatial variables are denoted by $^+$ when normalized by the viscous length scale $\nu/u_\tau$, e.g. $y^+ = Re_\tau y$.

The velocity field can be represented by a weighted sum of resolvent modes.
The Fourier decomposition of the velocity field in the homogeneous directions $x$, $z$, and $t$ yields
\begin{equation}
	\bu (x,y,z,t)
	\; = \;
	\ds{
		\iiint_{-\infty}^{\infty}
	}
	\,
	\hat{\bu} (y, \kappa_x, \kappa_z, \omega)\,
	\mre^{\mri
		(
		\kappa_x x
		\, + \,
		\kappa_z z
		\, - \,
		\omega t
		)
	}
	\mrd \kappa_x \,
	\mrd \kappa_z \,
	\mrd \omega,
	\label{eq.fourier}
\end{equation}
where $\lambda_x$, $\lambda_z$, and $\omega$ denote the streamwise and spanwise wavelengths and the temporal frequency. The Fourier coefficients, denoted by $\hat{ }$ , are three-dimensional three-component propagating waves with streamwise and spanwise wavenumbers $\kappa_x = 2\pi/\lambda_x$ and $\kappa_z = 2\pi/\lambda_z$ and streamwise speed $c = \omega/\kappa_x$. The zero-wavenumber zero-frequency component is identified as the spatio-temporal mean ($\kappa_x = \kappa_z = \omega = 0$) $\bU = [\,U(y)~0~0\,]^T = \hat{\bu}(y,0,0,0)$ and the velocity fluctuations satisfy
\begin{equation}
	-\mri \omega \hat{\bu}
	\, + \,
	(\bU \cdot \nabla) \hat{\bu}
	\, + \,
	(\hat{\bu} \cdot \nabla) \bU
	\, + \,
	\nabla \hat{p}
	\, - \,
	(1/{Re}_\tau) \Delta \hat{\bu}
	\; = \;
	\hat{\fvec},
	~~
	\nabla \cdot \hat{\bu}
	\; = \;
	0,
	\label{eq.NS-lin}
\end{equation}
where $\fvec = [\,f_1~f_2~f_3\,]^T = -(\bu \cdot \nabla) \bu$ is considered as a forcing term that drives the fluctuations, $p$ is the pressure fluctuation, $\nabla = [\,\mri \kappa_x~\p_y~\mri \kappa_z\,]^T$, and $\Delta = \p_{yy} - \kappa^2$ where $\kappa^2 = \kappa_x^2 + \kappa_z^2$. The relationship between the nonlinear forcing and the velocity is described by
\begin{equation}
	\hat{\bu} (y, \blambda, c)
	\; = \;
	H (\blambda, c) \, \hat{\fvec} (y, \blambda, c),
	\non
\end{equation}
where $H$ is the resolvent operator and $\blambda = [\,\lambda_x~\lambda_z\,]$ is the wavelength vector. In the above and the rest of this paper, the variables are parameterised with $c$ instead of $\omega$ since $c$ plays an integral role in determining the appropriate scaling of the resolvent modes~\cite{moashatromckJFM13}. Notice that for given $\kappa_x$, knowledge of either $c$ or $\omega$ yields the other parameter.

Using the velocity-vorticity formulation to enforce the continuity equation, the resolvent operator is given by $H = C R B$ where
\begin{equation}
	\begin{array}
		{rcl}
		C
		  & \!\! = \!\! &
		\dfrac{1}{\kappa^2}
		\,
		\thbt
		{\mri \kappa_x \p_y}{-\mri \kappa_z}
		{\kappa^2}{0}
		{\mri \kappa_z \p_y}{\mri \kappa_x},
		~~
		B
		\; = \;
		\tbth
		{-\mri \kappa_x \Delta^{-1} \p_y}
		{\kappa^2 \Delta^{-1}}
		{-\mri \kappa_z \Delta^{-1} \p_y}
		{\mri \kappa_z}
		{0}
		{-\mri \kappa_x},
		\\[0.5cm]
		R
		  & \!\! = \!\! &
		\tbt
		{
		\Delta^{-1}
		\left(
		\mri \kappa_x \,
		( (U - {c}) \Delta \, - \, U'')
		\, - \,
		(1/{Re}_\tau) \Delta^2
		\right)
		}
		{0}
		{\mri \kappa_z U'}
		{
		\mri \kappa_x (U - {c})
		\, - \,
		(1/{Re}_\tau) \Delta
		}^{-1},
	\end{array}
	\non
\end{equation}
and $\Delta^2 = \p_{yyyy} - 2 \kappa^2 \p_{yy} + \kappa^4$ and prime denotes differentiation in $y$, e.g. $U' (y) = \mrd U/\mrd y$.

For any $\blambda$ and $c$, the Schmidt (singular value) decomposition of $H$ in the non-homogeneous direction $y$ yields an orthonormal set of forcing modes $\hat{\bphi}_j = [\,\hat{f}_{1j}~\hat{f}_{2j}~\hat{f}_{3j}\,]^T$ and an orthonormal set of response (resolvent) modes $\hat{\bpsi}_j = [\,\hat{u}_j~\hat{v}_j~\hat{w}_j\,]^T$ that are ordered by the corresponding gains $\sigma_1 \geq \sigma_2 \geq \cdots \geq 0\,$ such that $H \hat{\bphi}_j = \sigma_j \hat{\bpsi}_j$. Therefore, if the nonlinear forcing is approximated by a weighted sum of the first $N$ forcing modes,
\begin{equation}
	\begin{array}
		{rcl}
		\hat{\fvec} (y, \blambda, c)
		  & \!\! = \!\! &
		\ds{\sum_{j = 1}^{N}}
		\;
		\chi_j (\blambda, c)\,
		\,
		\hat{\bphi}_j (y, \blambda, c),
	\end{array}
	\label{eq.f}
\end{equation}
the velocity is determined by a weighted sum of the first $N$ resolvent modes,
\begin{equation}
	\begin{array}
		{rcl}
		\hat{\bu} (y, \blambda, c)
		  & \!\! = \!\! &
		\ds{\sum_{j = 1}^{N}}
		\;
		\chi_j (\blambda, c)\, \sigma_j (\blambda, c)\,
		\,
		\hat{\bpsi}_j (y, \blambda, c).
	\end{array}
	\label{eq.u}
\end{equation}
The complex weights $\chi_j$ may be obtained by projecting the nonlinear forcing onto the forcing modes,
\begin{equation}
	\begin{array}
		{rcl}
		\chi_j (\blambda, c)
		  & \!\! = \!\! &
		\ds{\int_{0}^{2}}
		\hat{\bphi}_j^* (y, \blambda, c)
		\, \cdot
		\hat{\fvec} (y, \blambda, c)
		\,
		\mrd y,
	\end{array}
	\label{eq.f-phi}
\end{equation}
where the star denotes the complex conjugate.
Expressing the NSE in terms of the mode coefficients results in a quadratic equation in the coefficients, which may then be solved.

Much of our work to date has focused on the form and scaling of the response and forcing modes, $\hat{\bpsi}_j (y, \blambda, c)$ and $\hat{\bphi}_j (y, \blambda, c)$, much of which is associated with the presence of a critical layer where $U(y)=c$, as summarised in \cite{mckshajac13} and \cite{moashatromckJFM13}. In the present work, we focus on the scaling of the nonlinear interaction term induced by the scaling of the mean velocity.
As a prerequisite, we revisit the known scaling results derived for the resolvent, $H$.

\subsection{Scaling of the resolvent induced by the mean velocity profile}
\label{sec.resolventscaling}

\begin{table}
	\centering
	\begin{tabular}{l|cccccc}
		Class
		\hspace{-0.05cm}
		  &
		\hspace{-0.05cm}
		range of $c$
		\hspace{0.05cm}
		  &
		\hspace{0.03cm}
		$\lambda_x$
		\hspace{0.03cm}
		  &
		\hspace{0.03cm}
		$y, \lambda_z$
		\hspace{0.03cm}
		  &
		\hspace{-0.03cm}
		$\sigma_j$
		\hspace{-0.03cm}
		  &
		\hspace{-0.15cm}
		$\hat{u}_j, \hat{f}_{2j}, \hat{f}_{3j}$
		\hspace{-0.15cm}
		  &
		\hspace{-0.15cm}
		$\hat{v}_j, \hat{w}_j, \hat{f}_{1j}$
		\hspace{-0.15cm}
		\\[0.2cm] \hline
		Inner
		\hspace{-0.05cm}
		  &
		\hspace{-0.05cm}
		$0 \leq c \leq 16$
		\hspace{0.05cm}
		  &
		\hspace{0.03cm}
		$Re_\tau^{-1}$
		\hspace{0.03cm}
		  &
		\hspace{0.03cm}
		$Re_\tau^{-1}$
		\hspace{0.03cm}
		  &
		\hspace{-0.03cm}
		$Re_\tau^{-1}$
		\hspace{-0.03cm}
		  &
		\hspace{-0.15cm}
		$Re_\tau^{1/2}$
		\hspace{-0.15cm}
		  &
		\hspace{-0.15cm}
		$Re_\tau^{1/2}$
		\hspace{-0.15cm}
		\\[0.2cm]
		Self-similar
		\hspace{-0.05cm}
		  &
		\hspace{-0.05cm}
		$16 \leq c \leq U_{cl} - 6.15$
		\hspace{0.05cm}
		  &
		\hspace{0.03cm}
		$y_c^+ y_c$
		\hspace{0.03cm}
		  &
		\hspace{0.03cm}
		$y_c$
		\hspace{0.03cm}
		  &
		\hspace{-0.03cm}
		$(y_c^+)^2 y_c$
		\hspace{-0.03cm}
		  &
		\hspace{-0.15cm}
		$y_c^{-1/2}$
		\hspace{-0.15cm}
		  &
		\hspace{-0.15cm}
		$(y_c^+)^{-1} y_c^{-1/2}$
		\hspace{-0.15cm}
		\\[0.2cm]
		Outer
		\hspace{-0.05cm}
		  &
		\hspace{-0.05cm}
		$0 \leq U_{cl} - c \leq 6.15$
		\hspace{0.05cm}
		  &
		\hspace{0.03cm}
		$Re_\tau$
		\hspace{0.03cm}
		  &
		\hspace{0.03cm}
		$1$
		\hspace{0.03cm}
		  &
		\hspace{-0.03cm}
		$Re_\tau^2$
		\hspace{-0.03cm}
		  &
		\hspace{-0.15cm}
		$1$
		\hspace{-0.15cm}
		  &
		\hspace{-0.15cm}
		$Re_\tau^{-1}$
		\hspace{-0.15cm}
	\end{tabular}
	\caption{Scaling for the inner, outer, and self-similar classes of the resolvent modes~\cite{moashatromckJFM13}. The range of mode speeds that distinguish these classes and the growth/decay rates (with respect to $Re_\tau$ or $y_c$) of the wall-parallel wavelengths, height, gain, and forcing and response modes are shown. The self-similar and outer scales are valid for the modes with aspect ratio $\lambda_x/\lambda_z \geq \gamma$, where a conservative value for $\gamma$ is $\sqrt{3}$ for the self-similar class and $\sqrt{3} Re_\tau$ for the outer-scaled class. The critical wall-normal location corresponding to the mode speed is denoted by $y_c$, i.e. $c = U (y_c)$.}
	\label{table.scalings}
\end{table}        		

The resolvent operator admits three classes of scaling on $(\blambda,c)$ and $y$ such that the appropriately-scaled resolvent modes are independent of $Re_\tau$ and, under certain conditions, also geometrically self-similar \cite{moashatromckJFM13}. This is summarised in table~\ref{table.scalings}. The scaling primarily depends on the mode speed and is associated with the different regions of the turbulent mean velocity. We have used the classical overlap layer representation of the mean velocity profile,
\begin{equation}
    {
    U(y^+) = B + \kappa^{-1} \ln (y^+)
    }
\end{equation}
but other forms can also be investigated. {Here, $B$ = 4.3 and the K{\'a}rm{\'a}n constant $\kappa$ = 0.39 optimally match the logarithmic region of the measured mean velocity in the mean-square sense~\cite{marmonhulsmi13}.}

The critical layer for a particular mode is defined as the wall-normal location $y_c$ where the mode's speed equals the local mean velocity, $c = U(y_c)$. This critical layer typically acts on the leading resolvent modes at that wavespeed, to localise them in the wall-normal direction, such that the peak streamwise velocity occurs at or near $y_c$~\cite{mcksha10, mckshajac13, shamck13}. Thus it becomes convenient to parameterise the wall-normal location of the mode centre with $y_c$. The appropriate scaling of modes in the wall-normal direction directly results from localisation of the resolvent modes around this critical layer. The scaling in the wall-parallel directions follows from the balance between viscous dissipation $(1/Re_\tau) (\mathrm{d}^2/\mathrm{d} y^2 - \kappa_x^2 - \kappa_z^2)$ and advection by the mean velocity $\mathrm{i} \kappa_x U$.

As summarised in table \ref{table.scalings}, in the inner scaling region of the mean velocity, the modes scale in inner units. We have taken $0 \le y^+ \le 100$, such that $0 \leq c \leq U(y^+ = 100) \approx 16$ as a representative range.  That is to say, mode shapes varying over $Re_\tau$ collapse for constant $(\lambda_x^+, \lambda_z^+)$ and constant $c$.

In the outer, wake region of the mean velocity profile, $0 \leq U_{cl} - c \leq U_{cl} - U(y = 0.1) =  6.15$ ($U_{cl} = U(y = 1)$ denotes the centreline velocity), the modes scale in outer units. Modes with intermediate wavespeed corresponding to the overlap layer of the mean velocity profile, $16 \leq c \leq U_{cl} - 6.15$ are geometrically self-similar, scaling with the distance  of their centre from the wall, $y_c$, where $c = U(y_c)$.

The modes in the self-similar and outer-scaled classes must satisfy an aspect-ratio constraint $\lambda_x/\lambda_z \geq \gamma$, where a conservative value for $\gamma$ is $\sqrt{3}$ for the self-similar class and $\sqrt{3} Re_\tau$ for the outer-scaled class.
As discussed in~\cite{moashatromckJFM13}, this is because the balance between $(1/Re_\tau) (\mathrm{d}^2/\mathrm{d} y^2 - \kappa_x^2 - \kappa_z^2)$ and $\mathrm{i} \kappa_x U$ for self-similar modes ($y \sim y_c$) requires that the viscous dissipation due to spanwise gradients is sufficiently larger than streamwise gradients. Here, we require that $\kappa_z^2$ is $3$ times larger than $\kappa_x^2$ or $\lambda_x/\lambda_z > \gamma = \sqrt{3}$. The balance between the above terms for outer-scaled modes ($y \sim 1$) requires $\gamma = \sqrt{3} Re_\tau$ using a similar argument. The selected aspect ratios are conservative because the dissipation due to wall-normal gradients $\mathrm{d}^2/\mathrm{d} y^2$ can dominate the dissipation due to $x$ and $z$ gradients when $\kappa_x$ and $\kappa_z$ are relatively small. In this case, the modes do not need to satisfy an aspect-ratio constraint. As a result, $\gamma$ depends on the second wall-normal derivative of the modes and finding a universal lower bound for $\gamma$ is difficult. However, since the energetic contribution of the modes with small spanwise wavenumbers is small, the selected aspect ratio is sufficient for the purpose of this paper.

The scaling of the streamwise component of the response modes was previously given in~\cite{moashatromckJFM13}. Here, we also report the scaling of the amplitudes of the wall-normal and spanwise response modes as well as all components of the forcing modes.

\subsection{Self-similar scaling and hierarchies in the log region}
\label{sec.hierarchies}

The scaling associated with the overlap region of the mean velocity admits
hierarchies of geometrically self-similar resolvent modes that are parameterised by $y_c$.
A hierarchy corresponds to a set of modes with constant $\lambda_x/(y_c^+ y_c)$ and $\lambda_z/y_c$, with $y_c$ located in the overlap region where the mean velocity can be represented as a logarithmic variation in $y$. We preserve generality by considering a logarithmic mean velocity for $y_l \leq y \leq y_u$ where $y_l$ and $y_u$ can admit a different scaling with $Re_\tau$, and we denote $c_l = U(y_l)$ and $c_u = U(y_u)$. The only \emph{a priori} bounds that are imposed on $y_l$ and $y_u$ correspond to the classical bounds for the top of the inner region and the bottom of the wake region in the mean velocity, i.e. $y_l^+ > 100$ and $y_u < 0.1$.

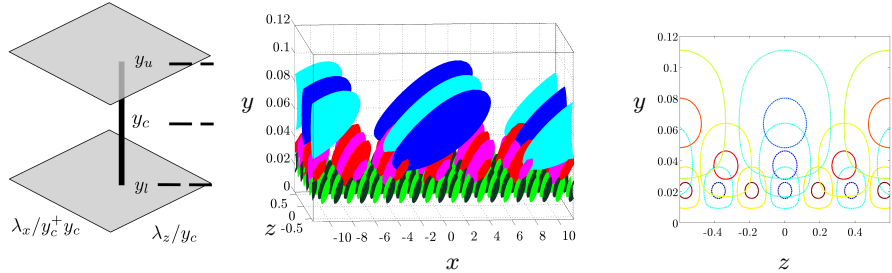
\begin{figure}
	\begin{center}
        \subfigure{
            \includegraphics[width=0.22\columnwidth]{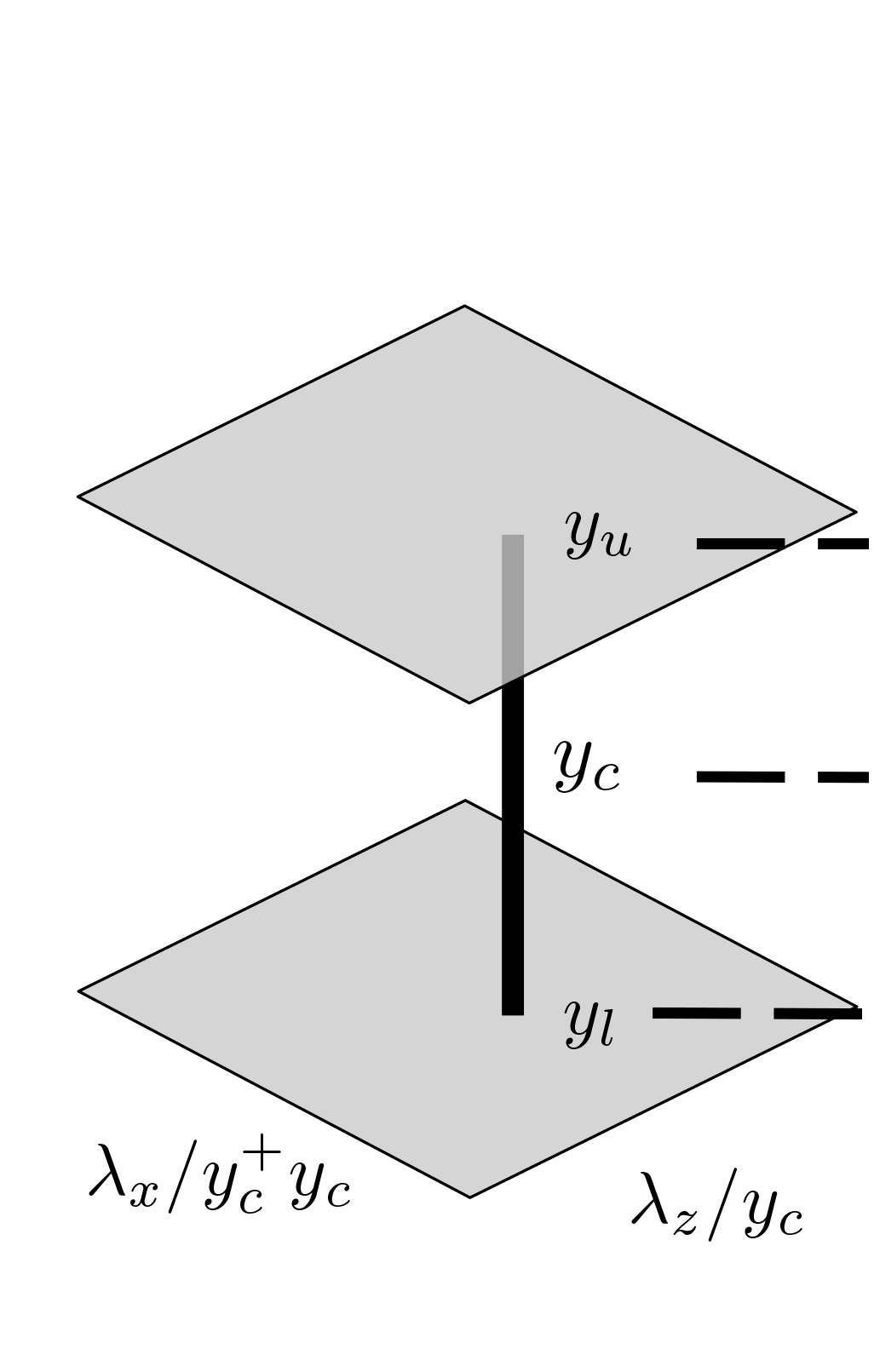}
            \label{fig.self-similar.schematic}
        }
        \subfigure{
            \begin{tikzpicture}
                \node[anchor=south west,inner sep=0] at (0,0) {
                    \includegraphics[width=0.35\columnwidth]{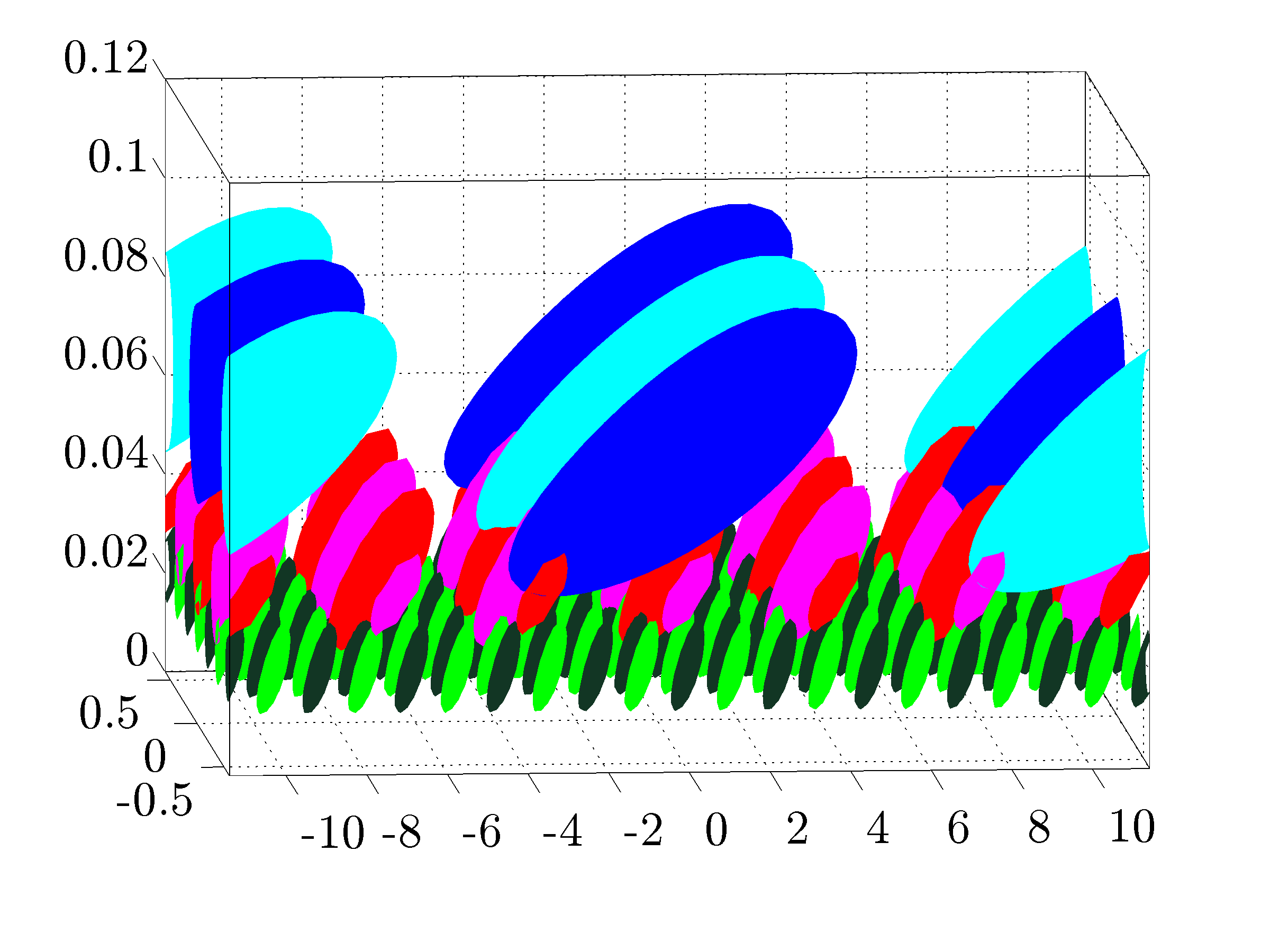}
                };
                \node at (2.7,0.1) {$x$};
                \node at (0,2.2) {$y$};
                \node at (0.3,0.55) {$z$};
            \end{tikzpicture}
            \label{fig.log-region-hierarchy-u-3D-R1e4-kx1-kz10-yc-0p0178-0p0316-0p0562}
        }
        \subfigure{
            \begin{tikzpicture}
                \node[anchor=south west,inner sep=0] at (0,0.3) {
                    \includegraphics[width=0.25\columnwidth]{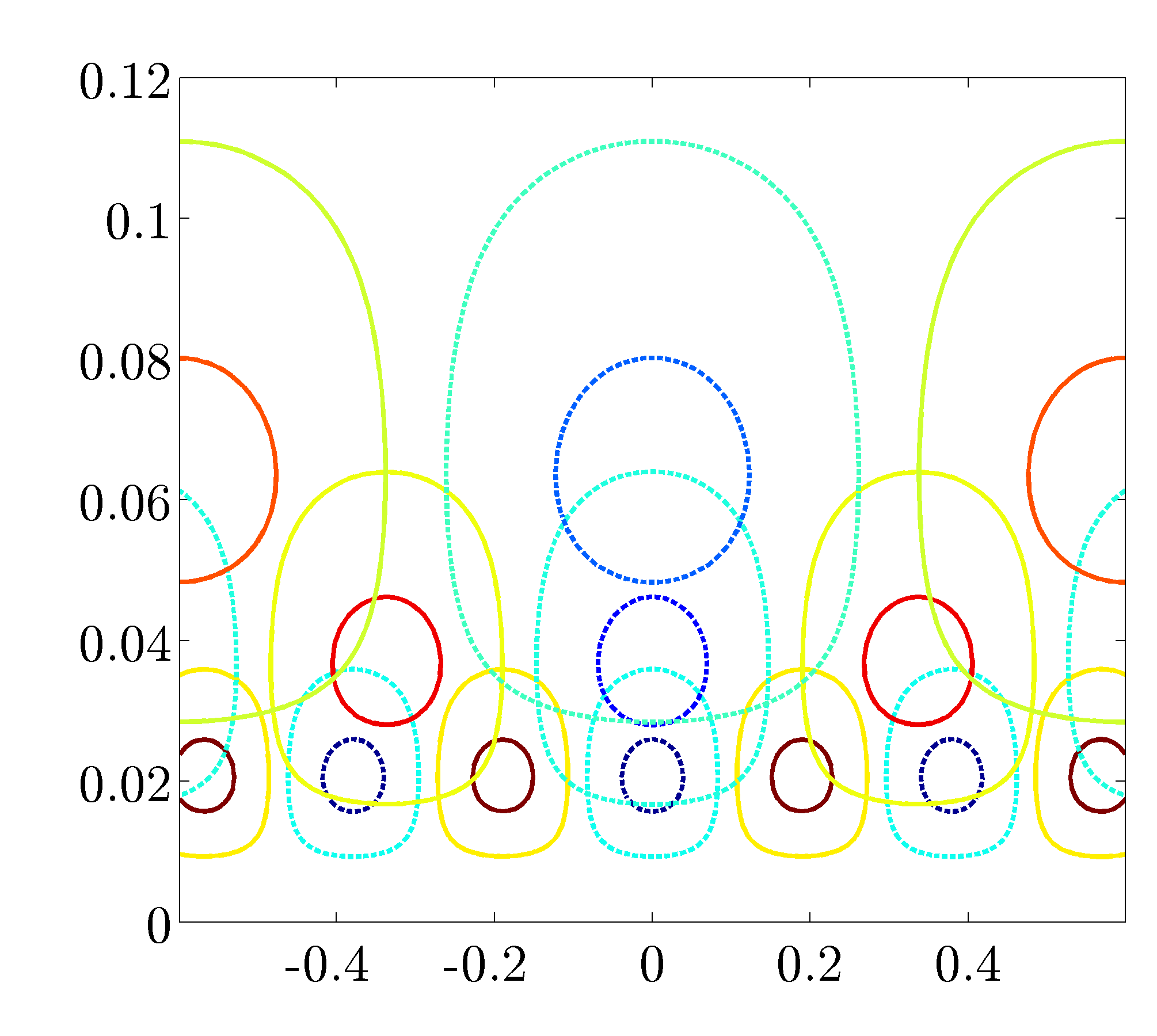}
                };
                \node at (1.9,0.1) {$z$};
                \node at (0,2.2) {$y$};
            \end{tikzpicture}
            \label{fig.log-region-hierarchy-contour-u-vs-lz-y-R1e4-kx1-kz10-yc-0p0178-0p0316-0p0562}
        }
	\end{center}
    \caption{(Left) Schematic showing that any mode in a given hierarchy (shown by the vertical line) is self-similar with respect to a reference mode in that hierarchy, and thus, can be expressed in terms of the reference mode.
    (Middle) Illustration of the geometrically self-similar resolvent modes: Isosurfaces of the principal streamwise velocities, $\hat{\bpsi}_1$, for three modes with $(\blambda, c) = (2.3,0.38,17.35)$; blue, $(7.2,0.67,18.70)$; red, and $(23,1.2,20.05)$; green, that belong to one hierarchy at $h^+ = 10^4$. The dark and light colours show $\pm70 \%$ of the maximum velocity. (Right) Cross-section of the middle plot at $z = 0$ showing contours of velocity at $\pm80 \%$ of the maximum.}
	\label{fig.self-similar}
\end{figure}

Figure~\ref{fig.self-similar.schematic} shows a schematic representation of the scaled wavenumber space in which any vertical line represents the locus of a hierarchy of self-similar resolvent modes. A mode may belong to one and only one hierarchy.

As $y_c$ or $c$ increases from $y_l$ to $y_u$, the modes become longer, taller and wider. It follows from the scaling of the wall-parallel wavelengths that the aspect ratio grows with $y_c^+$ within a hierarchy.

Isosurfaces of streamwise velocity associated with three modes that belong to single hierarchy are shown in figure~\ref{fig.log-region-hierarchy-u-3D-R1e4-kx1-kz10-yc-0p0178-0p0316-0p0562}. The larger modes propagate faster and lean more towards the wall since the length of the modes grows quadratically with the height. The cross-section of the streamwise velocity at $z = 0$ is shown in figure~\ref{fig.log-region-hierarchy-contour-u-vs-lz-y-R1e4-kx1-kz10-yc-0p0178-0p0316-0p0562}. As $c$ increases, the modes become larger and their centres move away from the wall. Specifically, their heights are proportional to the distance of their centres from the wall and their widths scale with their height.

The wavelengths and speed of the largest mode in a hierarchy can be denoted by $\blambda_u$ and $c_u$, with $\lambda_{x,u}/\lambda_{z,u} \geq \gamma$. Similarly, we denote the wavelengths and speed of the smallest mode in a hierarchy by $\blambda_{l'}$ and $c_{l'}$ where $\lambda_{x,l'} = (y_{l'}^+ y_{l'} / y_u^+ y_u) \lambda_{x,u}$, $\lambda_{z,l'} = (y_{l'} / y_u) \lambda_{z,u}$, and $c_{l'} = U(y_{l'})$. Here, $y_{l'}^+$ is the larger of the lower edge of the log region $y_l^+$ and the height $\gamma y_u^+ (\lambda_{z,u}/\lambda_{x,u})$ of the smallest mode that satisfies the aspect-ratio constraint. Therefore, the range of scales in a hierarchy depends on the ratio between $y_u$ and $y_{l'}$.

Because the modes are geometrically self-similar, any hierarchy of modes is characterized by the wavelengths $\blambda_r$ of an arbitrary reference mode in that hierarchy.  Formally, a hierarchy can then be defined as a subset $\cS (\blambda_r)$ of all mode parameters $\cS$ such that,
\begin{equation}
	\begin{array}
		{l}
		\cS (\blambda_r)
		\, = \,
		\Bigg\{
		(\blambda,c)
		\,|\,
		\lambda_x = \lambda_{x,r} \Big(\dfrac{y_c^+ y_c}{y_r^+ y_r}\Big),\
		\lambda_z = \lambda_{z,r} \Big(\dfrac{y_c}{y_r}\Big),\
		c = c_r + \dfrac{1}{\kappa} \ln \Big( \dfrac{y_c^+}{y_r^+} \Big),
		                                                                                        \\[0.3cm]
		\hskip2.85cm
		y_c^+ \geq y_{l'}^+ = \mbox{max} \{y_l^+, \gamma y_u^+ (\lambda_{z,u}/\lambda_{x,u})\},
		~y_c \leq y_u
		\Bigg\}.
	\end{array}
	\label{eq.S_h}
\end{equation}

Similarly, hierarchies at one Reynolds number can be determined from those at a reference Reynolds number. The inner-scaled variables $y_c^+$ and $y_r^+$ can be defined in terms of a reference Reynolds number for which the largest resolvent mode has been computed, $Re_{\tau,r}$, and the Reynolds number of interest, $Re_\tau$.
Then,
\begin{equation}
	y_c^+
	\, = \,
	Re_\tau y_c,~~
	y_r^+
	\, = \, Re_{\tau,r} y_r,
	\label{eq.ycp-yup}
\end{equation}
and substitution into (\ref{eq.S_h}) reveals that the characteristics of the hierarchies at arbitrary values of the Reynolds number are determined from those at the reference $Re_{\tau,r}$.

The mode shapes and their amplification can also be determined from the modes whose speed corresponds to the reference mode. Specifically, we have
\begin{equation}
	\begin{array}
		{rcl}
		g_1 (y, \blambda, c)
		  & \!\! = \!\! &
		\sqrt{y_r/y_c}
		\;
		g_1 \big(
		(y_r/y_c) y, \blambda_r, c_r
		\big),
		\\[0.2cm]
		g_2 (y, \blambda, c)
		  & \!\! = \!\! &
		(y_r^+/y_c^+) \sqrt{y_b/y_c}
		\;
		g_2 \big(
		(y_r/y_c) y, \blambda_r, c_r
		\big),
	\end{array}
	\label{eq.u-map-cu}
\end{equation}
where $g_1$ represents $\hat{u}_j$, $\hat{f}_{2j}$, or $\hat{f}_{3j}$ and $g_2$ represents $\hat{v}_j$, $\hat{w}_{j}$, or $\hat{f}_{1j}$. The corresponding singular values are obtained from
\begin{equation}
	\sigma_j (\blambda, c)
	\; = \;
	(y_c^+/y_r^+)^2 (y_c/y_r)
	\;
	\sigma_j (
	\blambda_r, c_r
	),
	\label{eq.sigma-map-cu}
\end{equation}
where we recall the distinction between $y_c^+$ and $y_r^+$, cf.~(\ref{eq.ycp-yup}).

Note that the aspect ratio $\lambda_x/\lambda_z$ decreases as $c$ becomes smaller in the log region. If a mode $m_0$ with speed $c_0$ belongs to a hierarchy, any mode with $c > c_0$ along the hierarchy also satisfies the aspect-ratio constraint and can be used to describe $m_0$. On the other hand, the modes with $c < c_0$ along the hierarchy may violate the aspect-ratio constraint, are excluded from the hierarchy, and cannot be used to retrieve $m_0$.

\section{Triadic interactions and self-similarity of the nonlinear interaction between modes}
\label{sec.scaling-weights}
\label{sec.triads}
\label{sec.coupling}

The development thus far has focused on the known scaling behaviour of the resolvent itself.  We now examine the implications of the geometrically self-similar scaling of the resolvent modes on the nonlinear interaction (coupling) between modes.

The quadratic nature of the nonlinearity in the NSE, as expressed by $\fvec$, implies that a resolvent mode with a given $(\blambda, c)$ can only be forced by pairs of modes that are triadically consistent, meaning that their streamwise wavenumbers, their spanwise wavenumbers and their temporal frequencies modes sum to give $(\blambda, c)$. It is clear that the modes' support must overlap in order for the corresponding forcing to be non-zero. Therefore triadic nonlinear interactions couple different scales in wavenumber-wavespeed space and different wall-normal locations in physical space. Previous work \cite{shamck13} explored the velocity field associated with a triadically consistent set of response modes; here we consider the characteristics of the forcing when the triad modes belong to geometrically self-similar hierarchies.

Following~\cite{mckshajac13}, an explicit equation for the weights which identifies the coupling between response modes in wavenumber/wavespeed space can be obtained. It follows from~(\ref{eq.f-phi}) that the weight $\chi_j (\blambda, c)$ is obtained by projecting the forcing $\hat{\fvec} (y, \blambda, c)$ onto the forcing mode $\hat{\bphi}_j (y, \blambda, c)$. Since $\fvec = -\bu \cdot \nabla \bu = -\nabla \cdot (\bu \bu^T)$, the Fourier-transformed forcing at a given $(\blambda, c)$ is given by the gradient of the convolution of all modes that are triadically-consistent with $(\blambda, c)$. The forcing associated with an individual triadic interaction is given by
\begin{equation}
	\begin{array}
		{l}
		{\hat{\fvec}}^* (y, \blambda, c)
		\; = \;
		-
		\nabla \cdot
		\,
		[\hat{\bu} (y, \blambda', c')
		\,
		\hat{\bu}^* (y, \blambda'', c'')
		\,+
\hat{\bu} (y, \blambda'', c'')
		\,
		\hat{\bu}^* (y, \blambda', c')
	\end{array}.
	\label{eq.f-triad}
\end{equation}
{
    The full forcing in physical space in terms of wavelengths, found by convolving all Fourier modes, is given by
}
\begin{equation}
	\begin{array}
		{l}
		{\hat{\fvec}}^* (y, \blambda, c)
						\; = \;
		-
		\nabla \cdot
				\ds{
		\iint
																				}
		\,
				\left(\dfrac{2\pi}{\lambda_x'}\right)^2 \dfrac{2\pi}{|\lambda_z'|}
		\,
								\hat{\bu} (y, \blambda', c')
		\,
		\hat{\bu}^* (y, \blambda'', c'')
		\,
		\mrd \ln \blambda' \,
		\mrd c'
				.
	\end{array}
	\label{eq.f-convolution}
\end{equation}
Here, we define for notational simplicity triadically-consistent wavelengths and wavespeed,
\begin{equation}
	\lambda_x'' \, = \, 	\dfrac{\lambda_x \lambda_x'}{\lambda_x' + \lambda_x},~~
	\lambda_z'' \, = \, \dfrac{\lambda_z \lambda_z'}{\lambda_z' + \lambda_z},~~
	c'' \, = \, \dfrac{c \lambda_x' + c' \lambda_x}{\lambda_x' + \lambda_x},
		\label{eq.mode-pp}
\end{equation}
and the symmetry relationships $\hat{\fvec} (y, -\blambda, c) = {\hat{\fvec}}^* (y, \blambda, c)$ and $\hat{\bu} (y, -\blambda, c) = {\hat{\bu}}^* (y, \blambda, c)$ are used. The mode speeds are confined to the interval $0 < c < U_{cl}$.

Substituting~(\ref{eq.f-convolution}) in~(\ref{eq.f-phi}) and using the symmetry relationship $\chi_i (-\blambda, c) = {\chi}^*_i (\blambda, c)$ gives the weight of the $l$-th response mode at $(\blambda,c)$,
\begin{equation}
	\begin{array}
		{l}
		{\chi}^*_l (\blambda, c)
		\; = \;
		\ds{
		\sum_{i,j = 1}^{N}
		\;
		\iint
																				}
		\,
		\cN_{lij} (\blambda,c,\blambda',c') \,
		\chi_i(\blambda', c') \,
		{\chi}^*_j(\blambda'', c'') \,
		\;
		\mrd \ln \blambda' \,
		\mrd c',
	\end{array}
	\label{eq.f-convolution-expand}
\end{equation}
where we have introduced the interaction coefficient, $\cN_{lij}(\blambda,c,\blambda',c')$, to describe the projection of the forcing arising from the interaction between two response modes onto the $l$-th forcing mode at $(\blambda, c)$, i.e.,
\begin{equation}
    \cN_{lij} (\blambda,c,\blambda',c') = - \sigma_i(\blambda', c') \sigma_j(\blambda'', c'') \int_0^2 \phi_{l}^*(y, \blambda, c) \cdot \left( \psi_i(y, \blambda', c') \cdot \nabla \psi_j(y, \blambda'', c'') \right) \,  dy.
\end{equation}

Expressed in this way, the interaction coefficient depends only on the coupling between (unweighted) resolvent modes and does not depend on the resolvent weights. This approach permits investigation of the nonlinear aspects without requiring knowledge of the weights corresponding to closing the system.  In this sense, the interaction coefficient provides a natural waypoint between the analysis of the linear resolvent operator and the full nonlinear system.

Notice that the expression for $\cN_{lij}$ is not symmetric with respect to swapping $i$ and $j$, i.e.~ in general $\cN_{lij} (\blambda,c,\blambda',c') \neq \cN_{lji} (\blambda,c,-\blambda'',c'')$, and one sense of interaction in a pair of resolvent modes may lead to a larger interaction coefficient than the other sense. By analogy to (\ref{eq.f-triad}) for the forcing, the total (symmetrised) coupling of $\hat{\bpsi}_i(\blambda',c')$ and $\hat{\bpsi}_j(-\blambda'',c'')$ to force $\hat{\bpsi}_l(-\blambda,c)$ is defined as
\begin{equation}
	\cN_{lij}^t (\blambda,c,\blambda',c')
	\; = \;
	\cN_{lij} (\blambda,c,\blambda',c')
	\, + \,
	\cN_{lji} (\blambda,c,-\blambda'',c'').
	\non
\end{equation}

Note also that, while we consider individual triads here, i.e. the forcing of an individual mode at $(\blambda,c)$, the statistical invariance in the wall-parallel directions and time implies the coexistence of a mode at $(-\blambda,-c)$ and supporting forcing. 
    
\subsection{Scaling of the interaction coefficient for the self-similar modes}
\label{sec.coupling-scaling}

\begin{figure}
	\begin{center}
		\begin{tabular}{cc}
			\subfigure{\includegraphics[width=0.41\columnwidth]{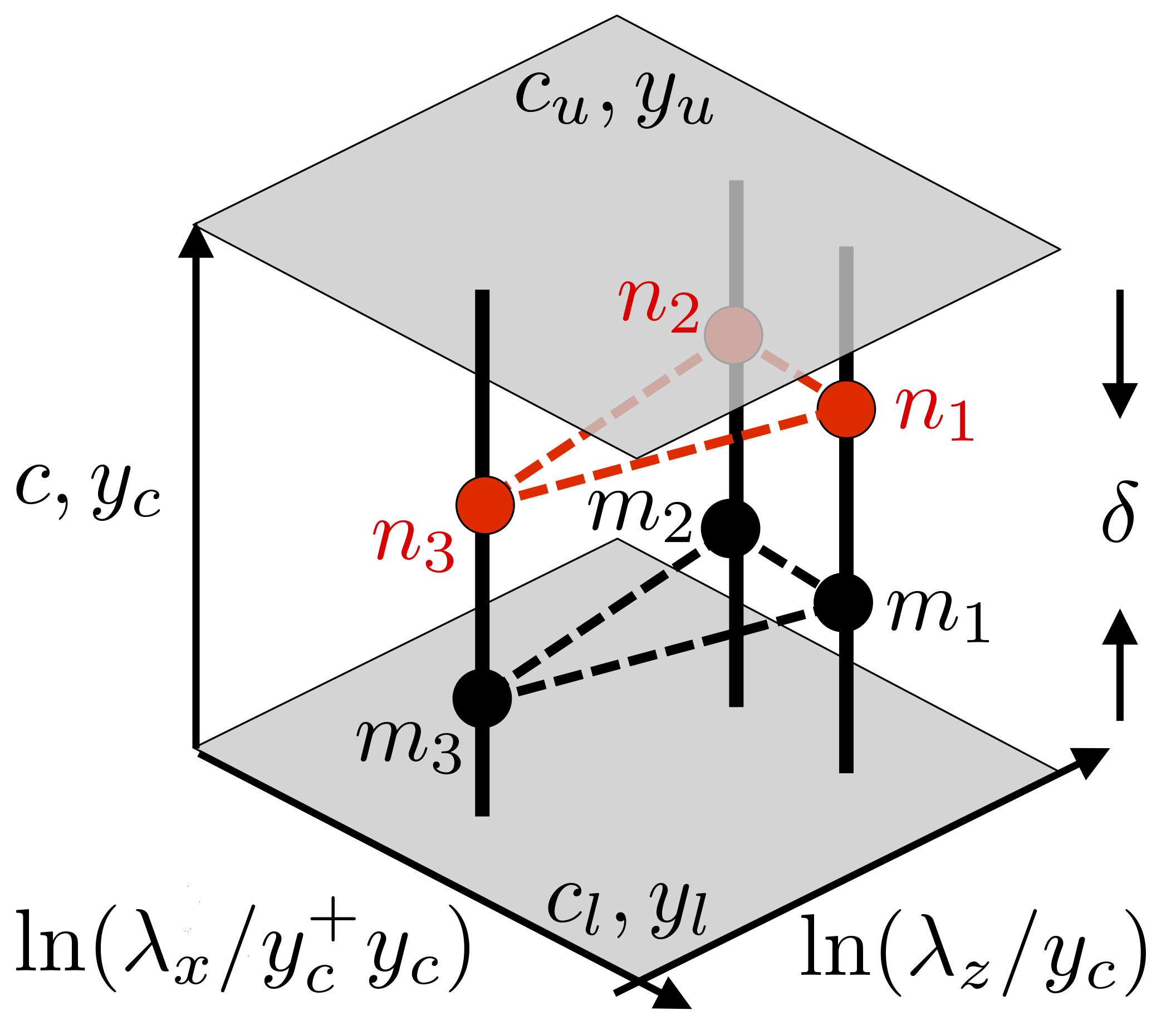}
			\label{fig.hierarchy-triad-1}}
			  &
			\subfigure{\includegraphics[width=0.45\columnwidth]{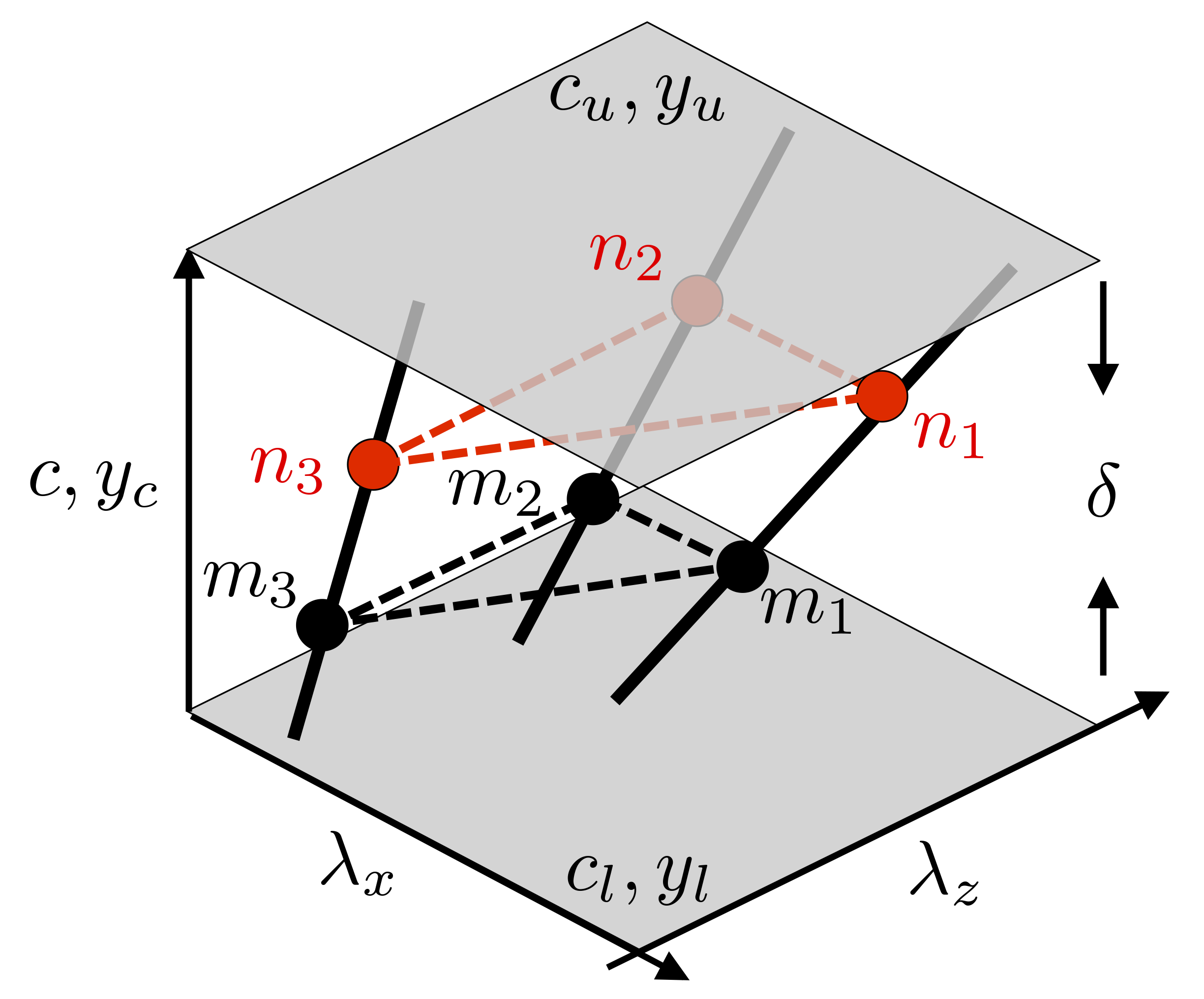}
			\label{fig.hierarchy-triad-alt}}
			\\[0.cm]
			$(a)$
			  &
			$(b)$
		\end{tabular}
	\end{center}
	\caption{Schematic showing triadically consistent self-similar hierarchies. The set of modes $m_1$, $m_2$, and $m_3$ are triadically consistent. The set of modes $n_1$, $n_2$, and $n_3$ are obtained by increasing the speeds of modes $m_1$, $m_2$, and $m_3$ along the corresponding hierarchies (vertical lines). As shown in table~\ref{table.triad}, the set of modes $n_1$, $n_2$, and $n_3$ are also triadically consistent. (a) Normalized wavelengths and (b) non-normalized wavelengths.}
	\label{fig.hierarchy-triad}
\end{figure}

The definition of self-similar hierarchies can be used to describe triadic interactions in the overlap region. Starting from any triad, moving an equal amount in $c$ along the hierarchies corresponding to the modes in that triad, we arrive at a new triad. This is illustrated in figure~\ref{fig.hierarchy-triad} and further explained in table~\ref{table.triad}, where the parameters for three triadically-consistent modes $m_1$, $m_2$, and $m_3$ are outlined. The corresponding modes $n_1$, $n_2$ and $n_3$ are obtained by moving along the hierarchies that include the modes $m_1$, $m_2$, and $m_3$ and increasing the mode wavespeeds by a constant $\delta = \kappa^{-1} \ln (\alpha^+)$. This increase moves the mode centres away from the wall (by $\alpha$ in outer units, $\alpha^+$ in inner units) and increases the mode wavelengths accordingly (shown in figure~\ref{fig.hierarchy-triad-alt}). The modes $n_1$, $n_2$ and $n_3$ are also triadically consistent and thus directly interact with each other.

\begin{figure}
	\begin{center}
        \subfigure{
            \includegraphics[width=0.54\columnwidth]{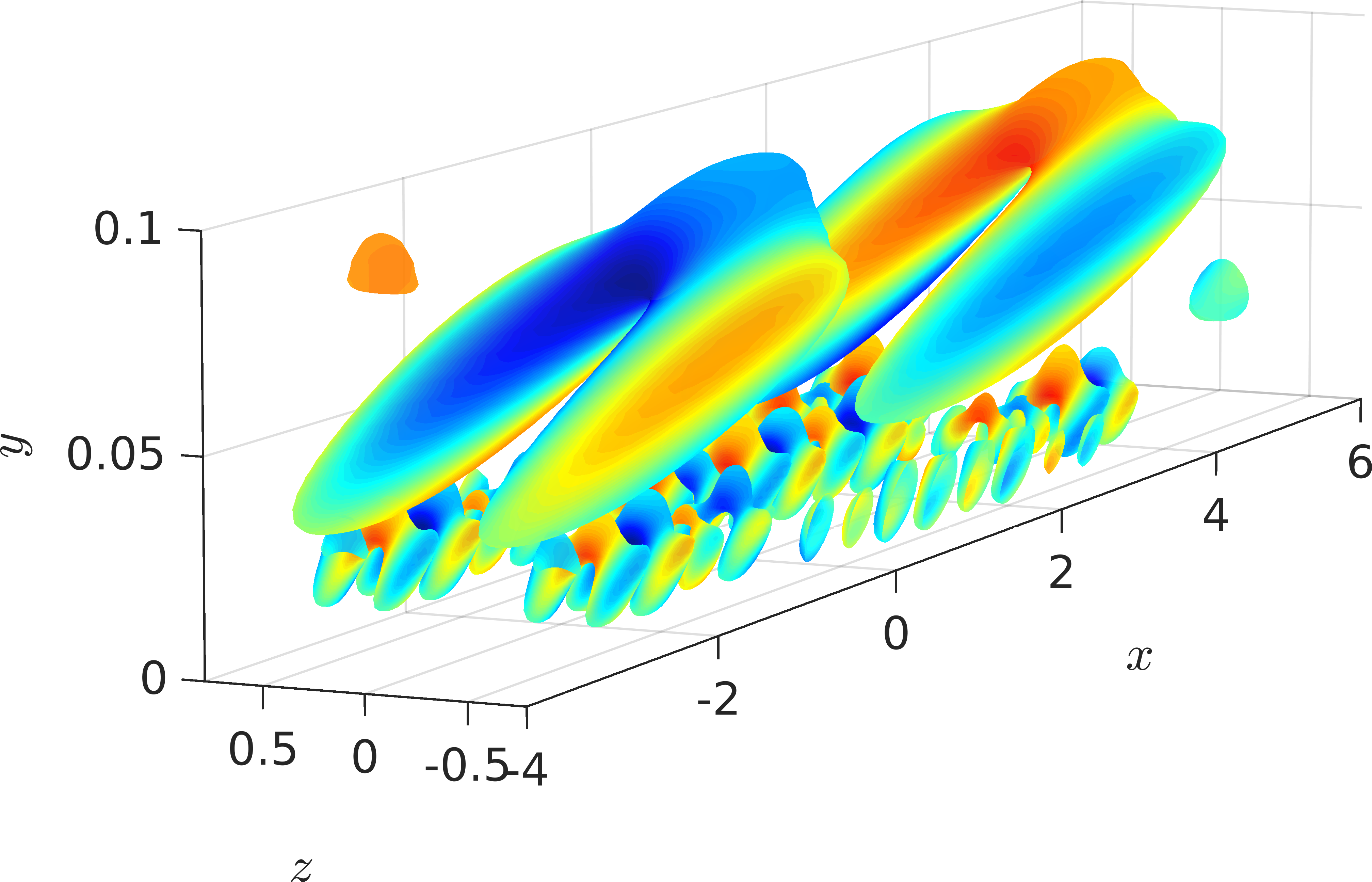}
            \label{fig.triad-swirl-2triads-3modes-3d-2016-07-11}
        }
        \hfill
		\subfigure{
            \includegraphics[width=0.32\columnwidth]{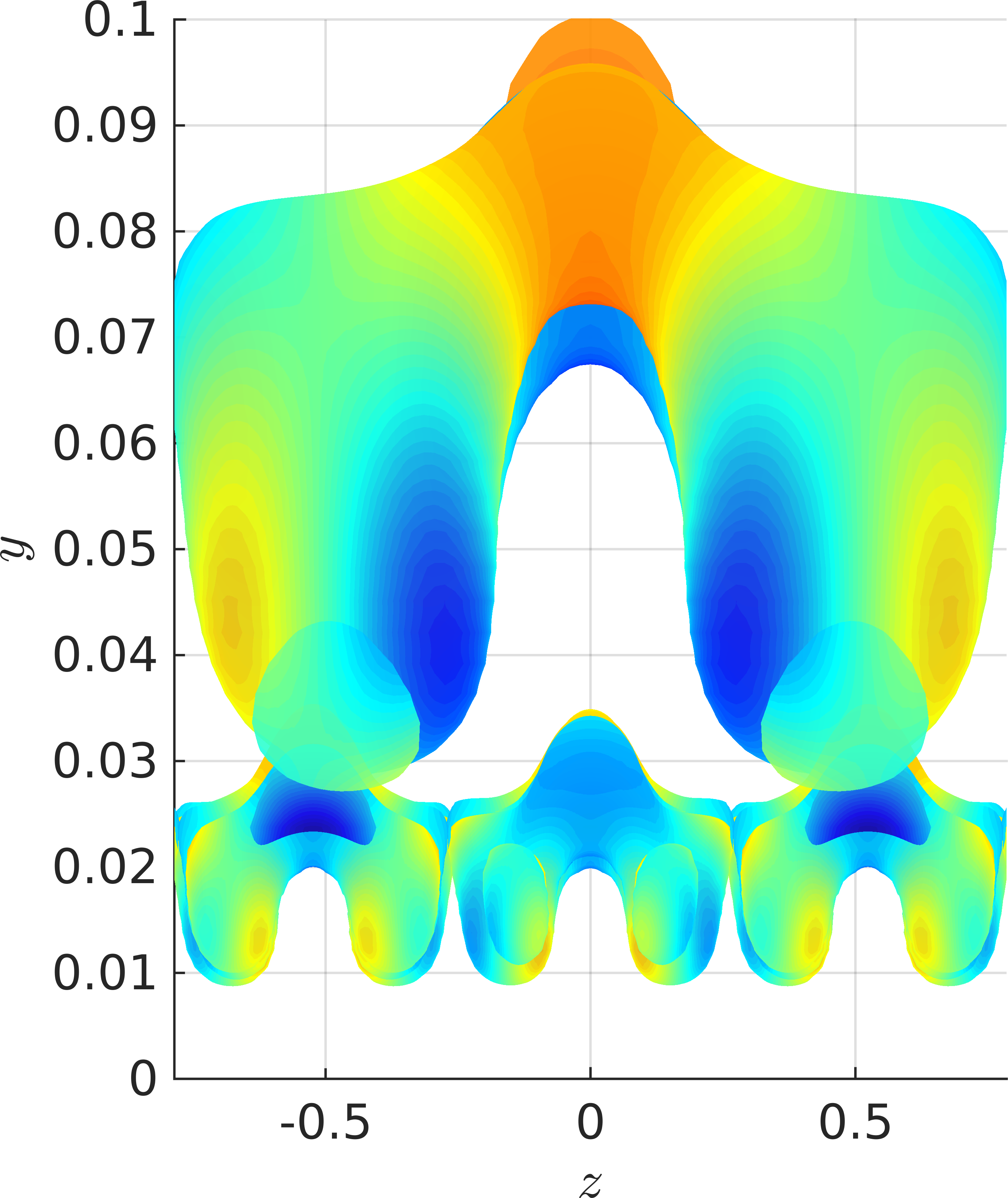}
			\label{fig.triad-swirl-2triads-3modes-yz-2016-07-11}
        }
	\end{center}
    \caption{The isosurfaces represent $50\%$ of the maximum swirling strength $\lambda_{ci}$ for two sets of triadically-consistent modes that belong to the same triadically-consistent hierarchies for $Re_\tau = 10^4$. The smaller/lower swirl structures respectively correspond to the triad modes $m_i$ with $(\blambda,c)_{m_1} = (2 \pi/6, 2\pi/6, 17)$, $(\blambda,c)_{m_2} = (2 \pi/1, 2\pi/6, 17)$ and $(\blambda,c)_{m_3} = (2 \pi/7, 2\pi/12, 17)$ and relative amplitudes $(0.05e^{-2.6i},0.25,0.045e^{-2.1i})$ after \cite{shamck13}. {The absolute phases differ from~\cite{shamck13} because here the phase gauge is defined such that the mode peaks at the $xz$-origin.} The larger/upper modes, $n_i$, are determined by the scaling in table~\ref{table.triad} with $\alpha^+=3$. The colours show the spanwise vorticity normalized by its maximum value where red (blue) denotes rotation in (opposite) the sense of the mean velocity. (Left) three-dimensional view and (right) cross-stream view.}
	\label{fig.swirl-triad}
\end{figure}

A turbulence ``kernel'' was previously proposed to capture important features of hairpin packet development and amplitude modulation behaviour \cite{shamck13}. The kernel used in that work was a triad of modes that included one representative of the very-large scale motion.
By way of illustration, figure~\ref{fig.swirl-triad} shows the swirl field associated with the sum of the velocity fields associated with both this kernel and the self-similar kernel obtained by moving upwards on the three hierarchies with $\alpha^+ = 3$. Consistent with the self-similar scaling with $y_c$, a geometrically self-similar array of hairpin-like vortices is observed.

\begin{table}
	\centering
	\begin{tabular}{c|cccc}
		mode
		  &
		$\lambda_x$
		  &
		$\lambda_z$
		  &
		$\omega$
		  &
		$c$
		\\[0.3cm] \hline
		$m_1$
		  &
		$\lambda_{x}$
		  &
		$\lambda_{z}$
		  &
		$\dfrac{2\pi c}{\lambda_{x}}$
		  &
		$c$
		\\[0.3cm]
		$m_2$
		  &
		$\lambda_{x}'$
		  &
		$\lambda_{z}'$
		  &
		$\dfrac{2\pi c'}{\lambda_{x}'}$
		  &
		$c'$
		\\[0.3cm]
		$m_3$
		  &
		\hskip0.3cm
		$-\dfrac{\lambda_{x} \lambda_{x}'}{\lambda_{x} + \lambda_{x}'}$
		\hskip0.3cm
		  &
		\hskip0.3cm
		$-\dfrac{\lambda_{z} \lambda_{z}'}{\lambda_{z} + \lambda_{z}'}$
		\hskip0.3cm
		  &
		\hskip0.3cm
		$-\dfrac{2\pi (c' \lambda_{x} + c \lambda_{x}')}{\lambda_{x} \lambda_{x}'}$
		\hskip0.3cm
		  &
		\hskip0.3cm
		$\dfrac{c' \lambda_{x} + c \lambda_{x}'}{\lambda_{x} + \lambda_{x}'}$
		\\[0.3cm]
		$n_1$
		  &
		$\alpha^+\alpha \lambda_{x}$
		  &
		$\alpha \lambda_{z}$
		  &
		$\dfrac{2\pi (c + \delta)}{\alpha^+\alpha \lambda_{x}}$
		  &
		$c + \delta$
		\\[0.3cm]
		$n_2$
		  &
		$\alpha^+\alpha \lambda_{x}'$
		  &
		$\alpha \lambda_{z}'$
		  &
		$\dfrac{2\pi (c' + \delta)}{\alpha^+\alpha \lambda_{x}'}$
		  &
		$c' + \delta$
		\\[0.3cm]
		$n_3$
		  &
		\hskip0.3cm
		$-\dfrac{\alpha^+\alpha \lambda_{x} \lambda_{x}'}{\lambda_{x} + \lambda_{x}'}$
		\hskip0.3cm
		  &
		\hskip0.3cm
		$-\dfrac{\alpha \lambda_{z} \lambda_{z}'}{\lambda_{z} + \lambda_{z}'}$
		\hskip0.3cm
		  &
		\hskip0.3cm
		$-\dfrac{2\pi ((c' + \delta) \lambda_{x} + (c + \delta) \lambda_{x}')}{\alpha^+\alpha \lambda_{x} \lambda_{x}'}$
		\hskip0.3cm
		  &
		\hskip0.3cm
		$\dfrac{c' \lambda_{x} + c \lambda_{x}'}{\lambda_{x} + \lambda_{x}'} + \delta$
	\end{tabular}	
    \caption{A set of triadically-consistent modes $m_1$, $m_2$, and $m_3$ and the set of modes $n_1$, $n_2$, and $n_3$ that are obtained by respectively moving along the hierarchies that include $m_1$, $m_2$, and $m_3$ such that the mode speeds increase with $\delta$. Relative to any of the modes $m_1$, $m_2$, and $m_3$, the centres of modes $n_1$, $n_2$, and $n_3$ move away from the wall by $\alpha$ in outer units and $\alpha^+$ in inner units where $\delta = \kappa^{-1} \ln (\alpha^+)$. Notice that $n_1$, $n_2$, and $n_3$ are triadically consistent themselves. See also figure~\ref{fig.hierarchy-triad}.
	}
	\label{table.triad}
\end{table}

The scaling of triadically-interacting hierarchies can be extended to consider the interaction coefficients associated with the self-similar modes. We consider the general case where the weights of the modes with speeds in the log region are primarily determined by the modes in the log region, so that all the interacting modes are self-similar. This is justified by the local interaction of the modes with each other as discussed earlier in~\S~\ref{sec.scaling-weights}.
\begin{figure}
	\begin{center}
		\begin{tabular}{cc}
			\subfigure{\includegraphics[width=0.42\columnwidth]{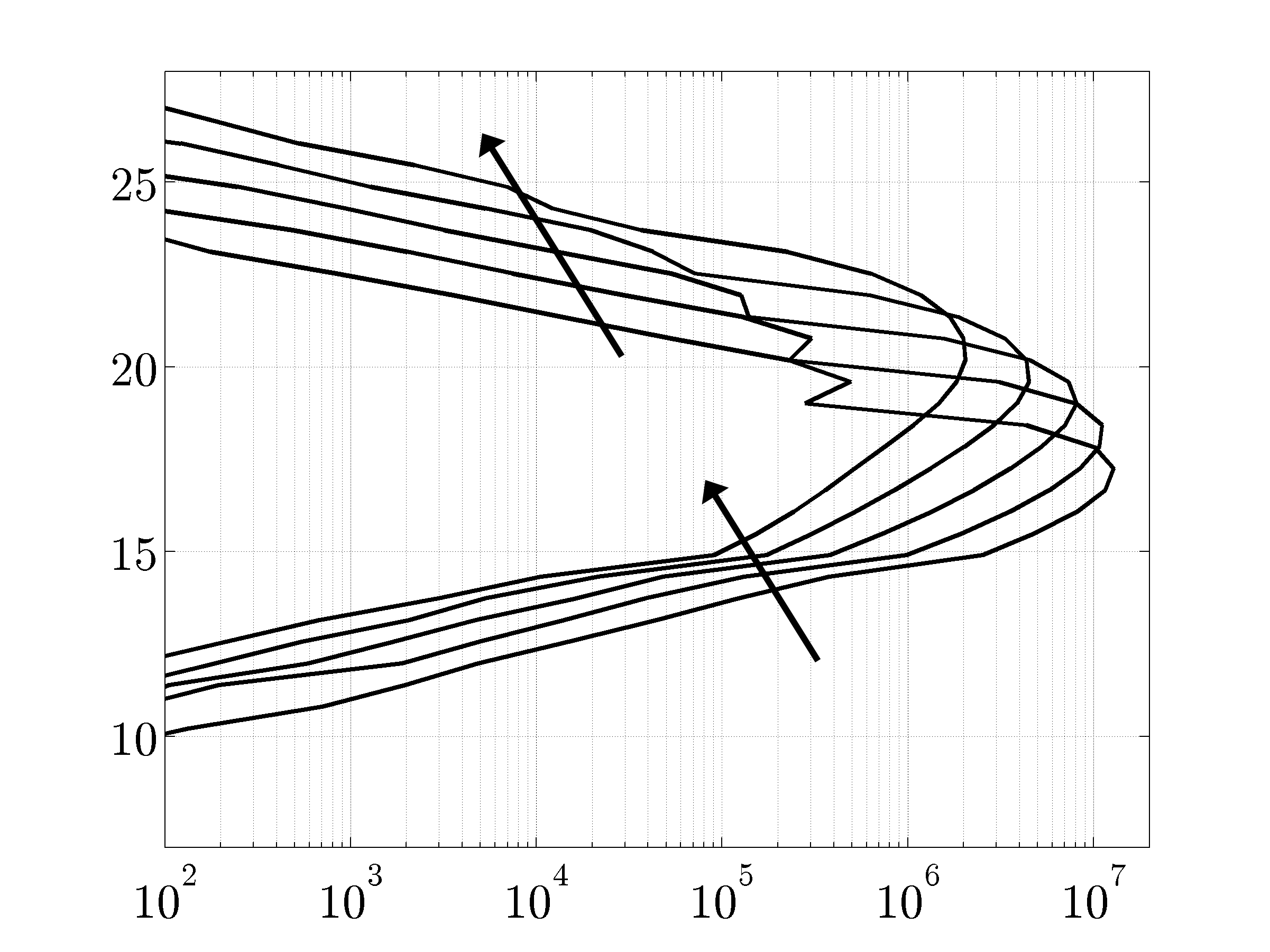}
			\label{fig.absNf-vs-cp2toUc-lx5p7-lz0p6-c23rdUc-lxpu0p35-lzpu0p11-R1e4-Q4-030414}}
			  &
			\subfigure{\includegraphics[width=0.42\columnwidth]{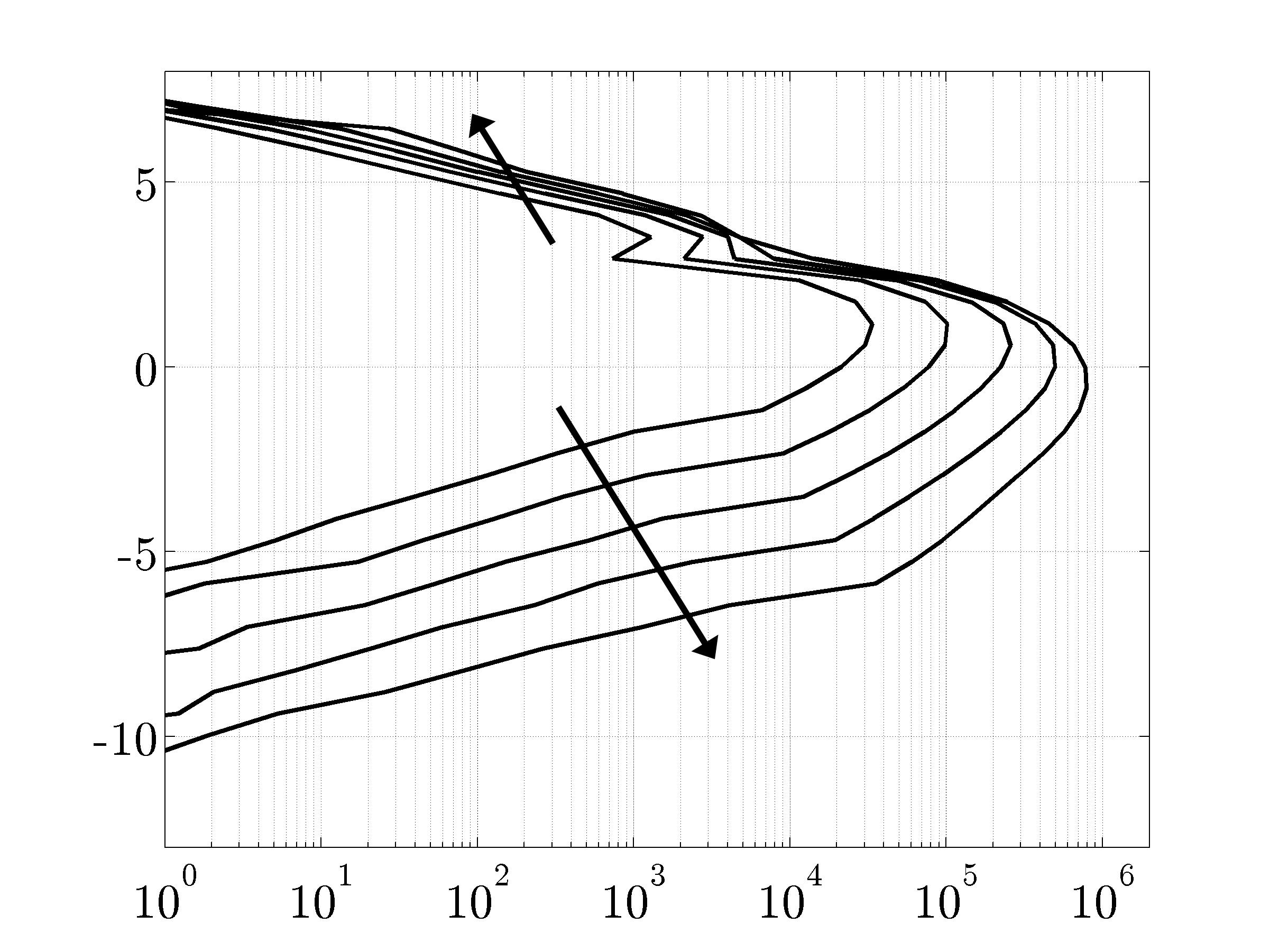}
			\label{fig.absNf-scaled-vs-cpmc1-lx5p7-lz0p6-c23rdUc-lxpu0p35-lzpu0p11-R1e4-Q4-030414}}
			\\[0.2cm]
			$(a)$
			  &
			$(b)$
		\end{tabular}
		\begin{tabular}{c}
			                     \\[-4.2cm]
			\begin{tabular}{c}
			\hskip-5.7cm
			\begin{turn}{90}
			\tc{black}{$~~c'$}
			\end{turn}
			\hskip5.7cm
			\begin{turn}{90}
			\tc{black}{$c' - c$}
			\end{turn}
		\end{tabular}
		\\[1.9cm]
		\begin{tabular}{c}
			\hskip0.1cm
			\tc{black}{$|\cN_{111}|$}
			\hskip5cm
			\tc{black}{$|\cM_{111}|$}
		\end{tabular}
		\end{tabular}
		\\[-0.2cm]
		\begin{tabular}{cc}
			\subfigure{\includegraphics[width=0.42\columnwidth]{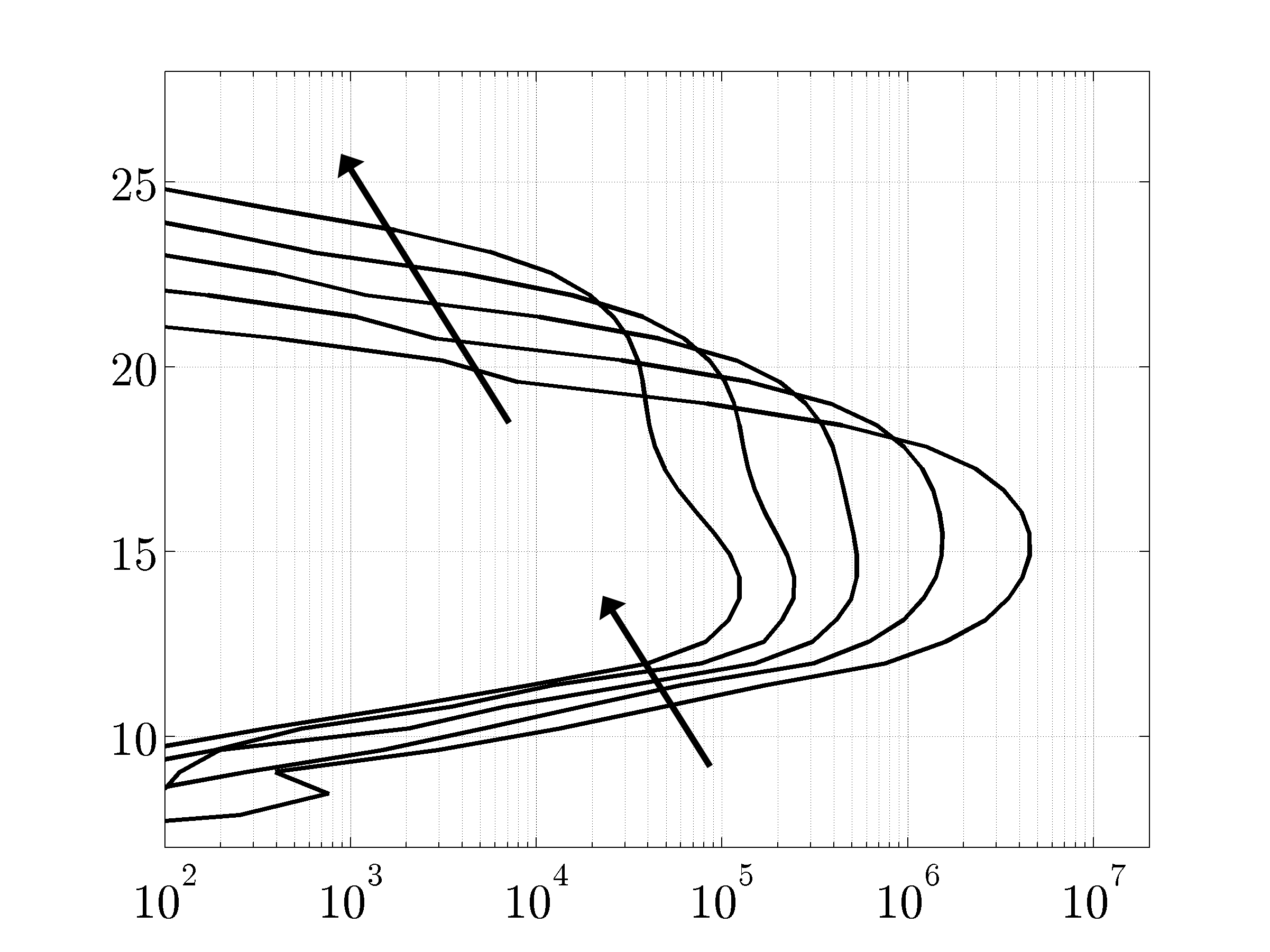}
			\label{fig.absNf-vs-cp2toUc-lx5p7-lz0p6-c23rdUc-lxpu1p11-lzpu0p035-R1e4-Q4-030414}}
			  &
			\subfigure{\includegraphics[width=0.42\columnwidth]{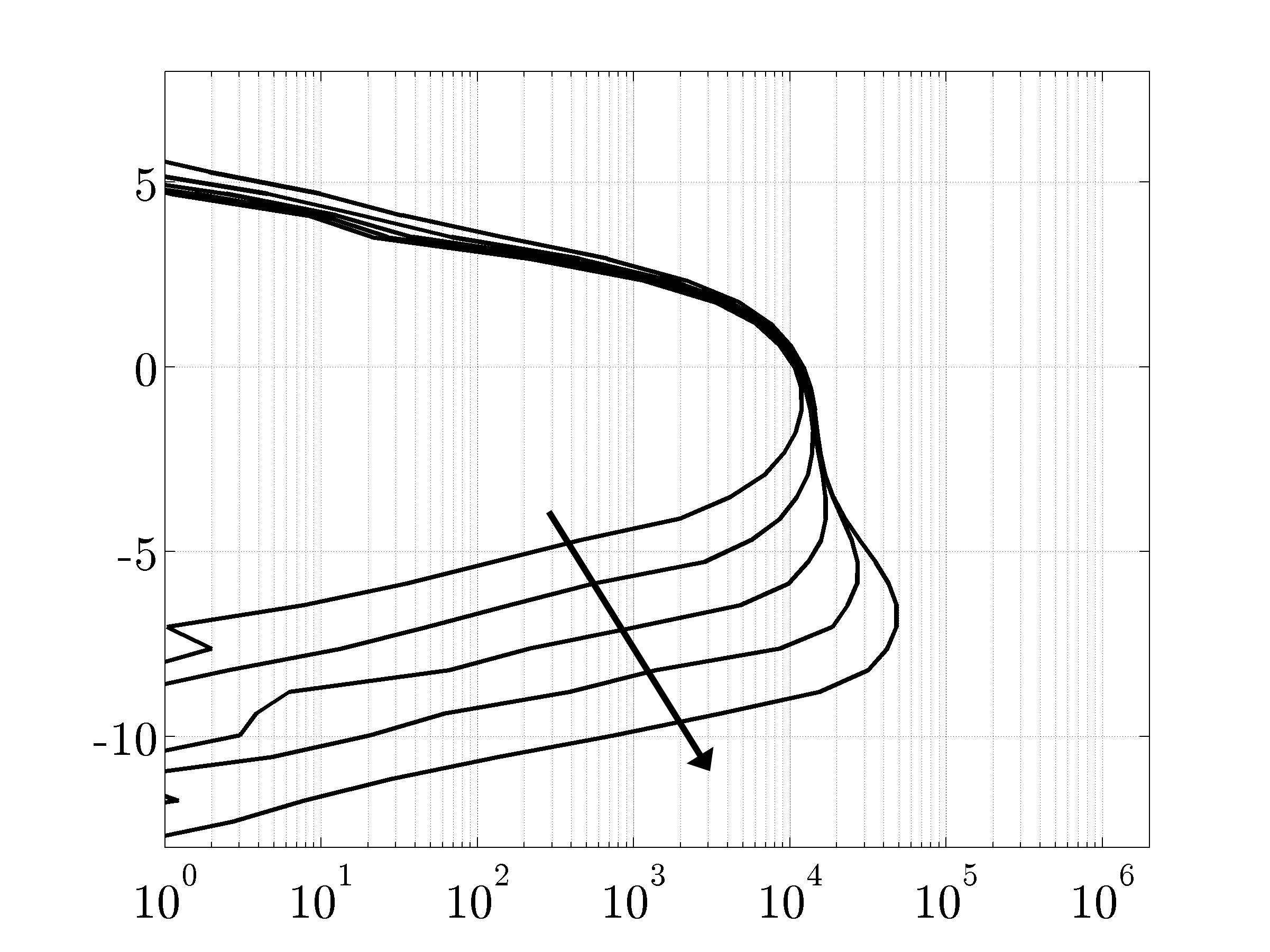}
			\label{fig.absNf-scaled-vs-cpmc1-lx5p7-lz0p6-c23rdUc-lxpu1p11-lzpu0p035-R1e4-Q4-030414}}
			\\[0.2cm]
			$(c)$
			  &
			$(d)$
		\end{tabular}
		\begin{tabular}{c}
			                     \\[-4.2cm]
			\begin{tabular}{c}
			\hskip-5.7cm
			\begin{turn}{90}
			\tc{black}{$~~c'$}
			\end{turn}
			\hskip5.7cm
			\begin{turn}{90}
			\tc{black}{$c' - c$}
			\end{turn}
		\end{tabular}
		\\[1.9cm]
		\begin{tabular}{c}
			\hskip0.1cm
			\tc{black}{$|\cN_{111}|$}
			\hskip5cm
			\tc{black}{$|\cM_{111}|$}
		\end{tabular}
		\end{tabular}
	\end{center}
	\caption{The absolute value of (a,c) the interaction coefficient $|\cN_{111} (\blambda,c,\blambda',c')|$ and (b,d) the self-similar interaction coefficient $|\cM_{111} (\blambda,\blambda',c'-c)|$ for $Re_\tau = 10^4$. Five forced modes $(\blambda, c)$ that belong to the hierarchy $h_1$ are considered. Arrows denote increasing $c$ on $h_1$. The forcing modes $(\blambda', c')$ belong to (a,b) the hierarchy $h_2$ with $\lambda_{x,u}' = 0.35$, $\lambda_{z,u}' = -0.11$ and (c,d) the hierarchy $h_3$ with $\lambda_{x,u}' = 1.11$, $\lambda_{z,u}' = -0.035$.}
	\label{fig.absN-scaled-vs-c1-cp-lx5p7-lz0p6-c23rdUc-R1e4}
\end{figure}

\begin{figure}
	\begin{center}
		\begin{tabular}{cc}
			\subfigure{\includegraphics[width=0.42\columnwidth]{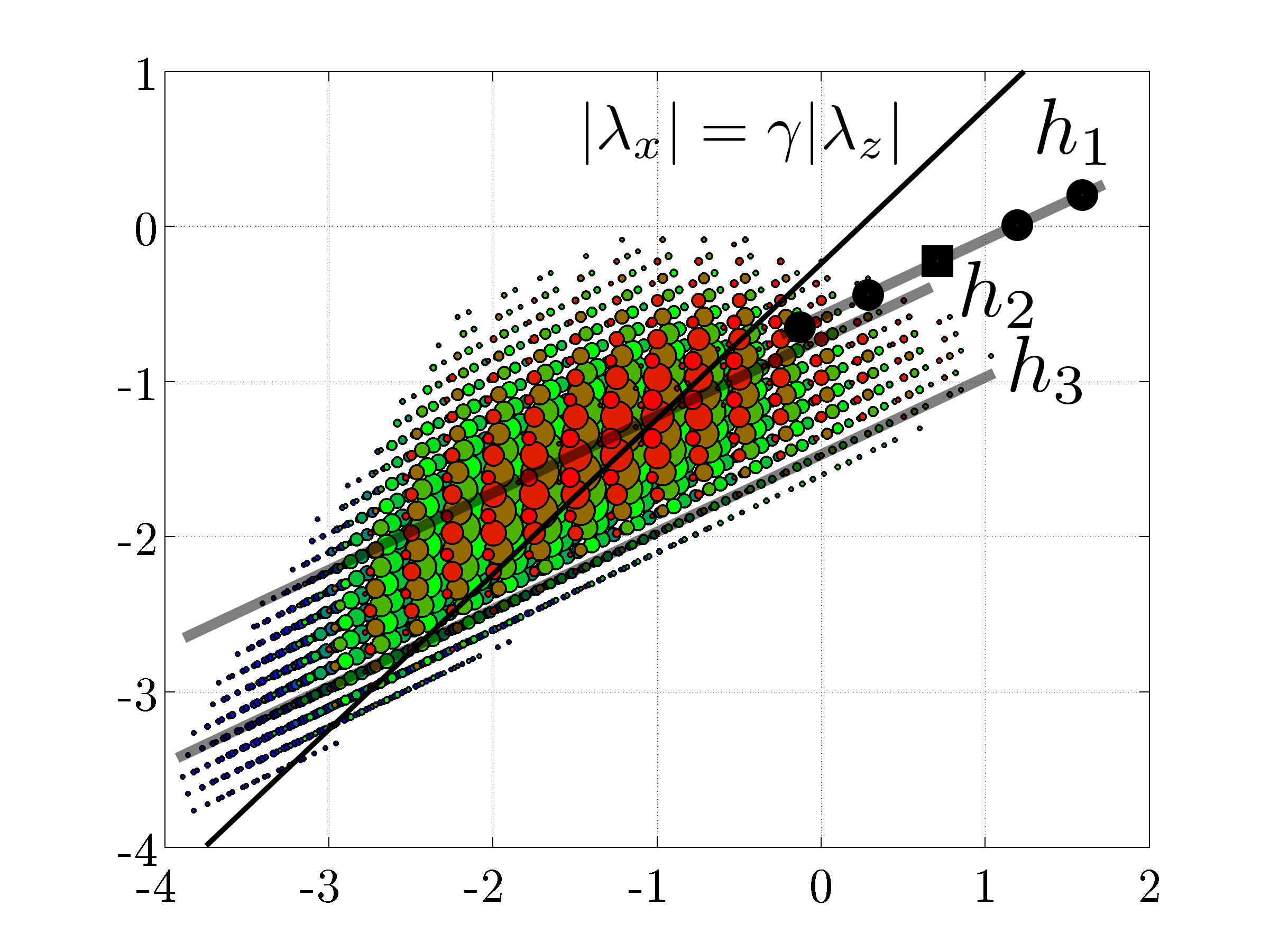}
			\label{fig.absN-vs-lxp-vs-lzp-lx5p7-lz0p6-c23rdUc-R1e4-Q4-h}}
			  &
			\subfigure{\includegraphics[width=0.42\columnwidth]{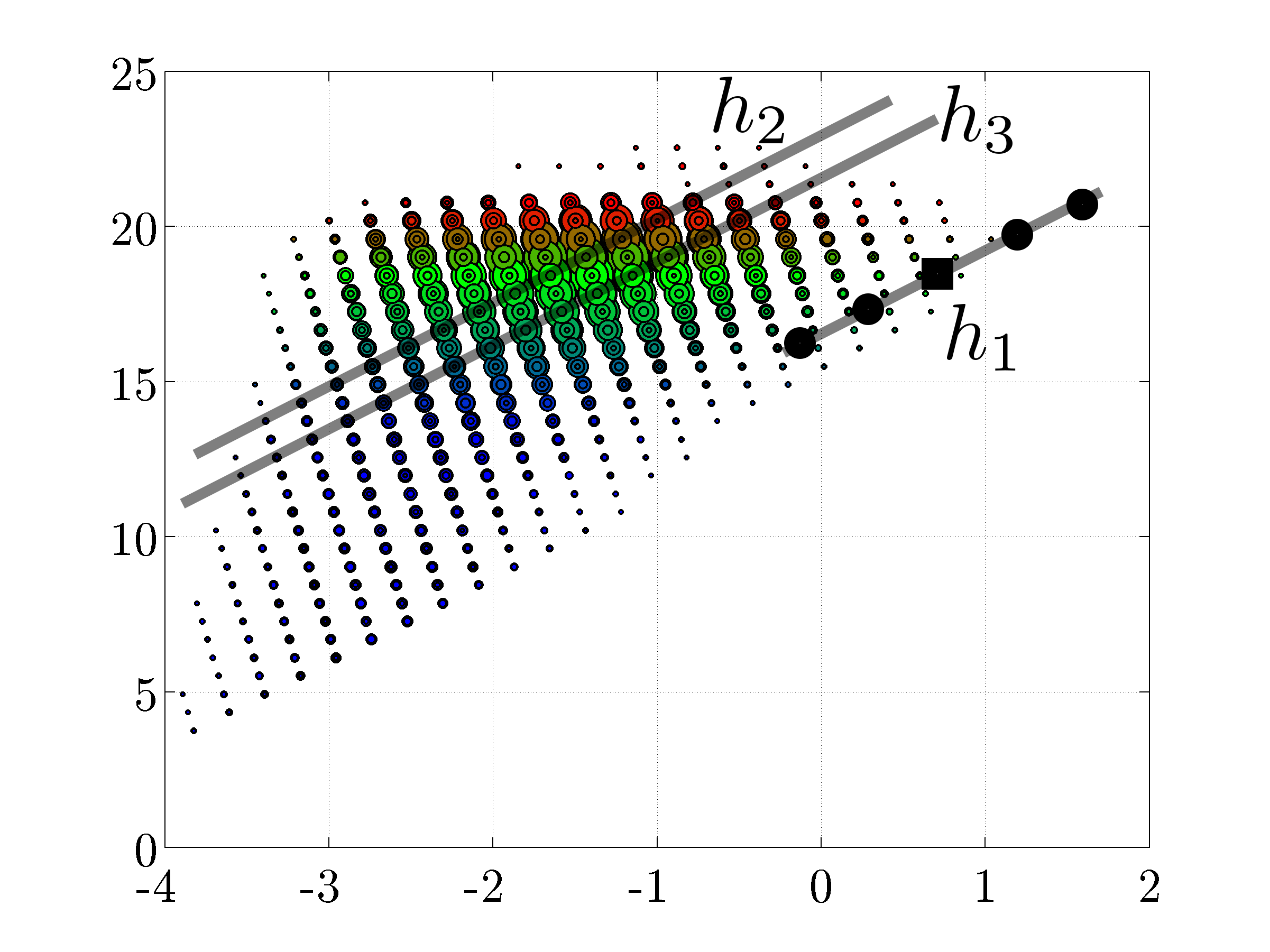}
			\label{fig.absN-vs-cp-vs-lxp-lx5p7-lz0p6-c23rdUc-R1e4-Q4-h}}
			\\[0.2cm]
			$(a)$
			  &
			$(b)$
		\end{tabular}
		\begin{tabular}{c}
			                                                  \\[-5cm]
			\begin{tabular}{c}
			\hskip-5.7cm
			\begin{turn}{90}
			\tc{black}{$\log |\lambda_z|, \log |\lambda_z'|$}
			\end{turn}
			\hskip5.7cm
			\begin{turn}{90}
			\tc{black}{$~~~~~~~~c, c'$}
			\end{turn}
		\end{tabular}
		\\[1.9cm]
		\begin{tabular}{c}
			\hskip0cm
			\tc{black}{$\log |\lambda_x|, \log |\lambda_x'|$}
			\hskip3.7cm
			\tc{black}{$\log |\lambda_x|, \log |\lambda_x'|$}
		\end{tabular}
		\end{tabular}
	\end{center}
    \caption{The absolute value of the interaction coefficient $|\cN_{111} (\blambda,c,\blambda',c')|$ for the representative VLSM mode with $\lambda_x = 5.7$, $\lambda_z = 0.6$, and $c = 18.4$, marked by the square, at $Re_\tau = 10^4$. The size of the coloured circles is proportional to $|\cN_{111}|$, and the circles are colour-coded by $c'$, plotted as a function of (a) $(\blambda')$; (b) $\lambda_x',c')$. {The largest and smallest circles correspond to $|\cN_{111}|=8.1 \times 10^6$ and $8.1\times 10^4$ respectively.} The diagonal black line in (a) denotes the aspect ratio for self-similarity. Also shown are the trajectories of the hierarchies $h_1-h_3$, where the VLSM (forced) mode sits on $h_1$. Circular black symbols on $h_1$ denote five forced modes $(\blambda, c)$ referenced in figure~\ref{fig.absN-scaled-vs-c1-cp-lx5p7-lz0p6-c23rdUc-R1e4} with mode speeds in the direction of the arrows are $c = 16$, $17.2$, $19.6$, and $20.8$.}
	\label{fig.absN-scaled-vs-c1-cp-lx5p7-lz0p6-c23rdUc-R1e4_Ns}
\end{figure}

The scaling of the resolvent modes (\ref{eq.u-map-cu}) and (\ref{eq.sigma-map-cu}), can be used to express (\ref{eq.f-convolution-expand}) in terms of the modes in the underlying hierarchies at a reference location:
Hereon, we use the wavelength of the upper mode in the hierarchy as the reference and assess the hierarchy based on the longest mode within it with $y_c$ chosen to be at the outer edge of the logarithmic region.

{
We will now present the derivation of the interaction coefficient scaling.
Substituting the nonlinear forcing term from~(\ref{eq.f-convolution}) in~(\ref{eq.f-phi}) yields~(\ref{eq.f-convolution-expand}) where
}
	\be
	\ba{l}
	\cN_{lij} (\blambda, c, \blambda', c')
	\; = \;
	-
	\big(\dfrac{2\pi}{\lambda_x'}\big)^2 \dfrac{2\pi}{|\lambda_z'|}
	\,
	\sigma_i(\blambda', c') 
	\,
	\sigma_j(\blambda'', c'') 
	\;
	\ds{
	\int_{0}^{2}
	}
	\Big\{
	\\[0.2cm]
	\hskip0.6cm
	\hat{f}_{1l} (y, \blambda, c)
	\Big(
	\big(
	\hat{u}_i (y, \blambda', c')
	\,
	\hat{v}^*_j (y, \blambda'', c'')
	\big)'
	\, + \,
	\\[0.2cm]
	\hskip1.8cm
	\mri 2\pi 
	\, 
	\hat{u}_i (y, \blambda', c')
	\,
	\big( 
	\hat{u}^*_j (y, \blambda'', c'')/\lambda_x  
	\, + \,
	\hat{w}^*_j (y, \blambda'', c'')/\lambda_z 
	\big)
	\Big)
	\, 
	+
	\\[0.2cm]
	\hskip0.6cm
	\hat{f}_{2l} (y, \blambda, c)
	\Big(
	\big(
	\hat{v}_i (y, \blambda', c')
	\,
	\hat{v}^*_j (y, \blambda'', c'')
	\big)'
	\, + \,
	\\[0.2cm]
	\hskip1.8cm
	\mri 2\pi 
	\, 
	\hat{v}_i (y, \blambda', c')
	\,
	\big( 
	\hat{u}^*_j (y, \blambda'', c'')/\lambda_x  
	\, + \,
	\hat{w}^*_j (y, \blambda'', c'')/\lambda_z 
	\big)
	\Big)
	\, 
	+
	\\[0.2cm]
	\hskip0.6cm
	\hat{f}_{3l} (y, \blambda, c)
	\Big(
	\big(
	\hat{w}_i (y, \blambda', c')
	\,
	\hat{v}^*_j (y, \blambda'', c'')
	\big)'
	\, + \,
	\\[0.2cm]
	\hskip1.8cm
	\mri 2\pi 
	\, 
	\hat{w}_i (y, \blambda', c')
	\,
	\big( 
	\hat{u}^*_j (y, \blambda'', c'')/\lambda_x  
	\, + \,
	\hat{w}^*_j (y, \blambda'', c'')/\lambda_z 
	\big)
	\Big)
	\,
	\Big\}
	\,
	\mrd y.
	\ea
	\label{eq.f-convolution-expand-ap}
	\ee

{
For a set of triadically-consistent modes in the self-similar hierarchies, notice that
}

\[
y_u/y_c \, = \, \mre^{\kappa (c_u - c)}, 
~~~
y_c/y_{c'} \, = \, \mre^{\kappa (c - c')}, 
~~~
y_c/y_{c''} \, = \, \mre^{\frac{\kappa \lambda_x}{\lambda_x + \lambda_x'} (c - c')}.
\]
{
Substituting the interaction coefficient from~(\ref{eq.f-convolution-expand-ap}) in~(\ref{eq.f-convolution-expand}) and defining
}
$
\tilde{y} \, = \, y y_u/y_c,
$
yields
\begin{equation}
	\begin{array}
		{l}
		{\chi}^*_l (\blambda, c)
		\; = \;
		\mre^{2.5 \kappa (c_u - c)}
		\;
		\ds{
		\sum_{i,j = 1}^{N}
		\;
		\iint
																				}
		\,
		\cM_{lij} (\blambda_u,\blambda_u',c'-c) \,
		\chi_i(\blambda', c') \,
		{\chi}^*_j(\blambda'', c'') \,
		\;
		\mrd \ln \blambda_u' \,
		\mrd c',
	\end{array}
	\label{eq.f-convolution-scale}
\end{equation}
where
	\be
	\ba{l}
	\cM_{lij} (\blambda_u, \blambda_u', c' - c)
	\; = \;
	\mre^{(3.5-1.5\frac{\lambda_x}{\lambda_x + \lambda_x'}) \kappa (c - c')}
	\,
	\big(\dfrac{2\pi}{\lambda_{x,u}'}\big)^2 \dfrac{2\pi}{|\lambda_{z,u}'|}
	\,
	\sigma_i(\blambda_{u}', c_u) \,
	\sigma_j(\blambda_{u}'', c_u) \,
	\;
	\ds{
      	\int_{0}^{2}
	}
	\Big\{
	\\[0.2cm]
	\hskip0.35cm
	\Big(
	\mre^{\kappa (c' - c)} 
	\, 
	\hat{f}_{1l} (\tilde{y}, \blambda_u, c_u)
	\,
	\big(
	\hat{u}_i (\tilde{y} \mre^{\kappa (c - c')}, \blambda_{u}', c_u)
	\,
	\hat{v}^*_{j} (\tilde{y} \mre^{\frac{\kappa \lambda_x}{\lambda_x + \lambda_x'} (c - c')}, \blambda_{u}'', c_u)
	\big)'
	\, + \,
	\\[0.2cm]
	\hskip1.76cm
	\hat{f}_{2l} (\tilde{y}, \blambda_u, c_u)
	\,
	\big(
	\hat{v}_i (\tilde{y} \mre^{\kappa (c - c')}, \blambda_{u}', c_u)
	\,
	\hat{v}^*_{j} (\tilde{y} \mre^{\frac{\kappa \lambda_x}{\lambda_x + \lambda_x'} (c - c')}, \blambda_{u}'', c_u)
	\big)'
	\, + \,
	\\[0.2cm]
	\hskip1.76cm
	\hat{f}_{3l} (\tilde{y}, \blambda_u, c_u)
	\,
	\big(
	\hat{w}_i (\tilde{y} \mre^{\kappa (c - c')}, \blambda_{u}, c_u)
	\,
	\hat{v}^*_{j} (\tilde{y} \mre^{\frac{\kappa \lambda_x}{\lambda_x + \lambda_x'} (c - c')}, \blambda_{u}'', c_u)
	\big)'
	\Big)
	\, + \,
	\\[0.2cm]
	\hskip0.35cm
	\mri 2\pi
	\,
	\Big(
	\,
	\mre^{\frac{\kappa \lambda_x}{\lambda_x + \lambda_x'} (c' - c)}
	\,
	\hat{u}^*_j (\tilde{y} \mre^{\frac{\kappa \lambda_x}{\lambda_x + \lambda_x'} (c - c')}, \blambda_{u}'', c_u)/\lambda_{x,u}
	\, + \,
	\hat{w}^*_j (\tilde{y} \mre^{\frac{\kappa \lambda_x}{\lambda_x + \lambda_x'} (c - c')}, \blambda_{u}'', c_u)/\lambda_{z,u}
	\Big)
	\Big(
	\\[0.2cm]
	\hskip1.2cm
	\mre^{\kappa (c' - c)}
	\,
	\hat{f}_{1l} (\tilde{y}, \blambda_u, c_u)
	\,
	\hat{u}_{i} (\tilde{y} \mre^{\kappa (c - c')}, \blambda_{u}', c_u)
	\, + \,
	\hat{f}_{2l} (\tilde{y}, \blambda_u, c_u)
	\,
	\hat{v}_{i} (\tilde{y} \mre^{\kappa (c - c')}, \blambda_{u}', c_u)
	\, + \,
	\\[0.2cm]
	\hskip1.2cm
	\hat{f}_{3l} (\tilde{y}, \blambda_u, c_u)
	\,
	\hat{w}_{i} (\tilde{y} \mre^{\kappa (c - c')}, \blambda_{u}', c_u)
	\Big)
	\,
	\Big\}
	\,
	\mrd \tilde{y}.
	\ea
	\label{eq.f-convolution-scale-ap}
	\ee
{
Notice that all the terms in~(\ref{eq.f-convolution-scale-ap}), including
}
\[
\dfrac{\lambda_x}{\lambda_x + \lambda_x'}
\, = \, 
\dfrac{1}{1 + \lambda_x'/\lambda_x}
\, = \,
\dfrac{1}{1 + (\lambda_{x,u}'/\lambda_{x,u}) \mre^{2 \kappa (c' - c)}},
\] 
can be expressed in terms of $\blambda_u$, $\blambda_u'$, and $c' - c$.

$\cM_{lij} (\blambda_u,\blambda_u',c'-c)$ is the ``self-similar interaction coefficient'' in the sense that for any modes $(\blambda, c) \in \cS(\blambda_u)$ and $(\blambda', c') \in \cS(\blambda_u')$, we have
\begin{equation}
	\cN_{lij} (\blambda,c,\blambda',c')
	\; = \;
	\mre^{2.5 \kappa (c_u - c)}
	\,
	\cM_{lij} (\blambda_u,\blambda_u',c'-c).
	\label{eq.M-N}
\end{equation}
Notice that $\cM$ only depends on the largest modes in the hierarchies that pass through the coupled modes. Therefore, the interaction coefficient for any set of triadically-consistent modes can be obtained from the interaction coefficient for the reference modes in the corresponding hierarchies. In other words, every interaction coefficient in the log region can be determined by the modes with speed $c_u = U(y_u)$. In addition, it follows from~(\ref{eq.M-N}) that within a set of triadically consistent hierarchies, the interaction coefficient is determined by the speed of the forced mode $c$ and the difference between $c$ and the speed of one of the forcing modes $c'$.

Figure~\ref{fig.absN-scaled-vs-c1-cp-lx5p7-lz0p6-c23rdUc-R1e4} shows the interaction coefficient for the five forced modes in the hierarchy $h_1$ identified by black filled symbols in figure~\ref{fig.absN-scaled-vs-c1-cp-lx5p7-lz0p6-c23rdUc-R1e4_Ns} and all the forcing modes in the hierarchies $h_2$ and $h_3$ marked by the shaded lines in figures~\ref{fig.absN-vs-lxp-vs-lzp-lx5p7-lz0p6-c23rdUc-R1e4-Q4-h} and~\ref{fig.absN-vs-cp-vs-lxp-lx5p7-lz0p6-c23rdUc-R1e4-Q4-h}.  Figures~\ref{fig.absNf-vs-cp2toUc-lx5p7-lz0p6-c23rdUc-lxpu0p35-lzpu0p11-R1e4-Q4-030414} and~\ref{fig.absNf-scaled-vs-cpmc1-lx5p7-lz0p6-c23rdUc-lxpu0p35-lzpu0p11-R1e4-Q4-030414} show $|\cN_{111} (\blambda,c,\blambda',c')|$ and $|\cM_{111} (\blambda,\blambda',c'-c)|$ for the forcing hierarchy $h_2$ with $\lambda_{x,u}' = 0.35$, $\lambda_{z,u}' = -0.11$. This hierarchy passes through the forcing modes that exhibit the largest interaction coefficient with the representative VLSM mode. As evident from figure~\ref{fig.absNf-vs-cp2toUc-lx5p7-lz0p6-c23rdUc-lxpu0p35-lzpu0p11-R1e4-Q4-030414}, $|\cN_{111}|$ peaks for $c' \approx c$ and decreases as $c$ becomes larger. Figure~\ref{fig.absNf-scaled-vs-cpmc1-lx5p7-lz0p6-c23rdUc-lxpu0p35-lzpu0p11-R1e4-Q4-030414} shows that the interaction coefficients are approximately self-similar for $2 < c' - c < 3$, notice the approximate collapse of $|\cM_{111}|$ in this region. For hierarchy $h_2$, the aspect-ratio constraint for self-similarity of the modes is satisfied only for large enough values of $c'$ (see figure~\ref{fig.absN-vs-lxp-vs-lzp-lx5p7-lz0p6-c23rdUc-R1e4-Q4-h}), leading to collapse only for a range of positive $c'-c$.

For comparison, we also consider the forcing hierarchy $h_3$ with $\lambda_{x,u}' = 1.11$, $\lambda_{z,u}' = -0.035$ where the aspect-ratio constraint is satisfied for a larger interval of $c'$ in the log region, cf. figure~\ref{fig.absN-vs-lxp-vs-lzp-lx5p7-lz0p6-c23rdUc-R1e4-Q4-h}. Figure~\ref{fig.absNf-vs-cp2toUc-lx5p7-lz0p6-c23rdUc-lxpu1p11-lzpu0p035-R1e4-Q4-030414} shows that $|\cN_{111} (\blambda,c,\blambda',c')|$ for $h_3$ locally peaks around $c' \approx c$ while a second peak emerges for $c' \approx 14$ as $c$ increases. Figure~\ref{fig.absNf-scaled-vs-cpmc1-lx5p7-lz0p6-c23rdUc-lxpu1p11-lzpu0p035-R1e4-Q4-030414} shows that the interaction coefficient is self-similar for $-1 < c' - c < 3$ and $c$ in the log region. Notice that the self-similarity extends to $|c' - c| < 3$ when only larger values of $c$ in the log region are considered. The self-similar interaction coefficients, at least for this triad, do not necessarily correspond to the largest ones, but we emphasize that the forcing is obtained by the product of the interaction coefficient and the weights corresponding to the forcing modes. Notice also the wide range in value of the interaction coefficient for varying $c'$ and hierarchy.

Figure~\ref{fig.absN-scaled-vs-c1-cp-lx5p7-lz0p6-c23rdUc-R1e4_Ns} shows the magnitude of the interaction coefficients, $|\cN_{111}|$, forcing a VLSM-like mode with $(\blambda,c)$ marked by the square symbol, specifically $\lambda_x = 5.7$, $\lambda_z = 0.6$, and $c = 18.4$ marked by the square in figures~\ref{fig.absN-vs-lxp-vs-lzp-lx5p7-lz0p6-c23rdUc-R1e4-Q4-h} and~\ref{fig.absN-vs-cp-vs-lxp-lx5p7-lz0p6-c23rdUc-R1e4-Q4-h}. Results are shown in terms of the $(\blambda',c')$ associated with one leg of the interacting modes. Clearly, the interaction coefficient is non-zero for a wide range of wavenumbers and wavespeeds, with large interactions confined to wavespeeds close to that of the VLSM, $c$.  Marked on these plots are the hierarchy to which the forced VLSM mode belongs, $h_1$, and two other hierarchies that will be investigated as part of a hierarchy of self-similar triads, $h_2-h_3$.

\section{Discussion and Conclusions}
\label{sec.conclusions}

We have seen that the resolvent operator admits a geometric self-similar scaling in the logarithmic region which is impressed on the basis functions (modes). When the nonlinear interaction of these modes is analysed, it is found that if three self-similar hierarchies are involved in a triadic interaction at one wavespeed, then they will also be triadically consistent after a constant increase in wavespeed on all hierarchies. The coefficient which describes their interaction also obeys a scaling.

As such, much information about the logarithmic region can be obtained by studying the resolvent operator and interaction coefficients at one reference wavespeed. The upper limit of the log region was selected here, but other choices are possible. In the long term, this may prove to have significant benefits for computational expense; scaling a mode is very much cheaper than calculating the singular value decomposition repeatedly for each position in the log layer.

It is perhaps significant that there are apparent differences with the scaling assumed by \cite{Woodcock.Marusic:2015} and found in simulation by \cite{Pirozzoli:2015}. 
In the equilibrium layer, Townsend used arguments concerning a dissipation length-scale proportional to distance from the wall to assume that the velocity field associated with the self-similar eddies is given by
\begin{equation}
	\bu (x,y,z)
	\; = \;
	s_1 \big(
	(x-x_a)/y_a, (y-y_a)/y_a, (z-z_a)/y_a
	\big),
	\non
\end{equation}
where $s_1$ is the velocity in terms of the normalized location of the eddy centre $x_a$, $y_a$, and $z_a$. The self-similar resolvent modes have the form
\begin{equation}
	\bu (x,y,z,t)
	\; = \;
	s_2 \big(
	(x - c\,t)/(y_c^+ y_c), (y-y_c)/y_c, z/y_c
	\big),
	\non
\end{equation}
where $s_2$ is the velocity in terms of the parameters that position the mode centres at the wall-parallel origin in a moving frame with streamwise speed $c$. 
The critical location $y_c$ in the present study is equivalent to Townsend's eddy centre $y_a$. In agreement with scaling of Townsend's eddies, the spanwise and wall-normal extents of the resolvent modes scale with $y_c$. On the other hand, the streamwise extent of the resolvent modes scales with $y_c^+ y_c$. Notice that this difference does not contradict Townsend's original hypothesis because the dissipation length-scale for the case where $\lambda_x$ and $\lambda_z$ are respectively proportional to $y_c^+ y_c$ and $y_c$ is dominated by the dissipation due to spanwise gradients, and hence, proportional to the mode height. It is intriguing that an analysis of the Navier-Stokes equations leads to such a discrepancy; a full description of the coefficient scaling would help to completely resolve this issue.

Note also that the scaling of the streamwise wavelength evokes the so-called ``mesolayer'' scaling, $y^{+}/\sqrt{Re_\tau} = \sqrt{y^+ y} = \mathrm{const}$, proposed as the inner limit of a logarithmic scaling region of both mean velocity and streamwise variance~\cite{marmonhulsmi13}. This scaling is also an integral part of the scale hierarchies that emerge from the mean momentum balance analysis, e.g. \cite{fifweiklemcm05}, suggesting a further significance to the present results that is yet to be determined.

The self-similarity of the resolvent and the interaction coefficients becomes less approximate, and governs a wider range of scales, as the Reynolds number increases. This suggests that it is directly amenable to exploitation at high Reynolds numbers. The scaling described herein is relevant to physical models, because it identifies the coupling between scales and wall-normal locations sustaining wall turbulence. It is also relevant to sub-grid-scale and wall models for large eddy simulation, where one objective is to understand and restrict the range of fully-resolved scales, augmenting them with models describing the unresolved scales.  The self-similarity in the log region seems ripe for exploitation in this sense.

While the system of equations still requires other methods to solve for the unknown coefficients and thereby become closed, the scaling derived herein is a necessary step to a complete description of the logarithmic layer within the framework. To our knowledge, it is the first observation of self-similarity in the nonlinear forcing. The next and final step in the development of a predictive model in the logarithmic region (and beyond) is to find a full description of how the coefficients scale. This is difficult, but is the subject of ongoing work.

\dataccess{For the purpose of EPSRC's data access policy, no significant data have been produced as part of this study.}

\aucontribute{
    RM carried out the original analysis.  All authors contributed to the manuscript and have approved it.
}

\competing{
    The authors declare that they have no competing interests.
}

\funding{
    The support of AFOSR grant FA9550-12-1-0469 and AFOSR/EOARD grant FA9550-14-1-0042 is gratefully acknowledged.
}

\bibliographystyle{ieeetr}
\bibliography{bib/couette,bib/mj-complete-bib,bib/periodic,bib/covariance,bib/control-pde,bib/ref-added-rm,bib/ati}

\end{document}